\documentclass[aps,prc,showpacs,showkeys,superscriptaddress,footinbib,eqsecnum]{revtex4}
\usepackage{graphicx}
\usepackage{amsmath}
\usepackage{amsfonts}
\usepackage{amssymb}
\usepackage{bm}
\newcommand{\beq}{\begin{equation}}
\newcommand{\eeq}{\end{equation}}
\newcommand{\beqa}{\begin{eqnarray}}
\newcommand{\eeqa}{\end{eqnarray}}
\newcommand{\be}{\begin{eqnarray}}
\newcommand{\ee}{\end{eqnarray}}
\newcommand{\eq}[1]{Eq.~(\ref{#1})}
\newcommand{\nn}{\nonumber \\ }

\date{\today}

\begin{document}

\title{High-quality two-nucleon potentials up to fifth order of the chiral expansion}

\author{D. R. Entem}
\email{entem@usal.es}
\affiliation{Grupo de F\'isica Nuclear, IUFFyM, Universidad de Salamanca, E-37008 Salamanca,
Spain}
\author{R. Machleidt}
\email{machleid@uidaho.edu}
\affiliation{Department of Physics, University of Idaho, Moscow, Idaho 83844, USA}
\author{Y. Nosyk}
\affiliation{Department of Physics, University of Idaho, Moscow, Idaho 83844, USA}

\begin{abstract}
We present $NN$ potentials through five orders of chiral effective field theory ranging from
leading order (LO) to next-to-next-to-next-to-next-to-leading order (N$^4$LO).
The construction may be perceived as consistent in the sense that the same power 
counting scheme as well as the same cutoff procedures are applied in all orders.
Moreover, the long-range parts of these potentials are fixed by the very accurate $\pi N$ LECs 
as determined in the Roy-Steiner equations analysis by Hoferichter, Ruiz de Elvira and coworkers. In fact,
the uncertainties of these LECs are so small that a variation within the errors leads to effects
that are essentially negligible, reducing the error budget of predictions considerably. The $NN$ potentials are fit to the world $NN$ data below pion-production threshold of the year of 2016.
 The potential of the highest order 
(N$^4$LO) reproduces the world $NN$ data with the outstanding
$\chi^2$/datum of 1.15, which is the highest precision ever accomplished for any chiral
$NN$ potential to date.
The $NN$ potentials presented may serve as a
solid basis for systematic {\it ab initio} calculations of nuclear structure and reactions
that allow for a comprehensive error analysis. In particular, the consistent order by order development of the potentials
will make possible a reliable determination of the truncation error at each order.
Our family of potentials is non-local and, generally, of soft character. This feature is reflected in the fact that
the predictions for the triton binding energy (from two-body forces only) converges to about 8.1 MeV
at the highest orders. This leaves room for three-nucleon-force contributions of moderate size.
\end{abstract}

\maketitle

\section{Introduction}
\label{sec_intro}
The quest for a practically feasible, and yet fundamental, theory of hadronic interactions at low energy (where QCD is non-perturbative) has spanned several decades. At the present time, there exists a general consensus that chiral effective field theory (chiral EFT) may provide the 
best answer to the quest. By its nature, chiral EFT is a model-independent approach with firm roots in QCD, due to the fact that interactions are subjected to the constraints of the broken chiral symmetry of low-energy QCD. Moreover, the approach is systematic in the sense that the various contributions to a particular dynamical process can be arranged as an expansion in terms of a suitable ``parameter''. The latter is chosen to be the ratio of a typical external momentum
(soft scale) to the chiral symmetry breaking scale ($\approx 1$ GeV, hard scale). Recent comprehensive reviews on the subject can be found in Refs.~\cite{ME11,EHM09}.   
      
  In its early stages, chiral perturbation theory (ChPT) was applied mostly to  $\pi\pi$~\cite{GL84} and $\pi N$~\cite{GSS88} dynamics, because, due to the Goldstone-boson nature of the pion,  these are the most natural scenarios for a perturbative expansion to exist. In the meantime, though, chiral EFT has been applied in nucleonic systems by numerous groups~\cite{ME11,EHM09,Wei90,ORK94,KBW97,KGW98,Kai00a,Kai00b,Kai01,Kai01a,Kai02,EGM98,EM02,EM03,EGM05,Nav07,Eks13,Gez14,Pia15,Pia16,EKM15,EKM15a,PAA15,Eks15,Car16,Tew16,Lyn16,Ren16}. Derivations of the nucleon-nucleon ($NN$) interaction up to fourth order (next-to-next-to-next-to-leading order, N$^3$LO) can be found in Refs.~\cite{KBW97,Kai00a,Kai00b,Kai01a,Kai02,EM02}, with quantitative $NN$ potentials making their appearance in the early 2000's~\cite{EM03,EGM05}.

Since then, a wealth of applications of N$^3$LO $NN$ potentials together with chiral three-nucleon forces (3NFs) have been reported. These investigations include few-nucleon 
reactions~\cite{Epe02,NRQ10,Viv13,Gol14}, structure of light- and medium-mass nuclei~\cite{BNV13,Her13,Hag14a,Sim17}, and infinite matter~\cite{HS10,Heb11,Hag14b,Cor13,Cor14,Sam15}.  Although satisfactory predictions have been obtained in many cases, persistent problems continue to pose serious challenges, such as the well-known `$A_y$ puzzle' of nucleon-deuteron scattering~\cite{EMW02}.
Naturally, one would invoke 3NFs as the most likely mechanism to solve this problem.
Unfortunately, the chiral 3NF at NNLO does only very little to improve the situation with nucleon-deuteron scattering~\cite{Epe02,Viv13}, while inclusion of the N$^3$LO 3NF produces an effect in the wrong direction~\cite{Gol14}. The next step is then to proceed systematically in the expansion, namely to look at N$^4$LO (or fifth order). 
This order is interesting for diverse reasons. From studies of some of the 3NF topologies at N$^4$LO~\cite{KGE12,KGE13}, we know that a complete set of isospin-spin-momentum 3NF structures (a total of 20) are present at this order~\cite{Epe15} and that contributions can be of substantial size. Even more promising, at this order a new set of 3NF contact interactions appears, which has recently been derived by the Pisa group~\cite{GKV11}. Contact terms are relatively easy to work with and, most importantly,  come with free coefficients and, thus, provide larger flexibility and a great likelihood to solve persistent problems such as 
 the $A_y$ puzzle as well as other issues (like, the ``radius problem''~\cite{Lap16} and the
 overbinding of intermediate-mass nuclei~\cite{Bin14}).
 
 A principle of all EFTs is that, for meaningful predictions, it is necessary 
 to include {\it all}\/ contributions that appear  
at the order at which the calculation is conducted. 
Thus, when nuclear structure problems require for their solution the inclusion of
3NFs at N$^4$LO, then also the two-nucleon force involved in the calculation has 
to be of order N$^4$LO. 
 This is one reason 
why in Ref.~\cite{Ent15a} we derived the N$^4$LO two-pion exchange (2PE) and 
three-pion exchange (3PE) contributions to the $NN$ interaction and tested them in peripheral partial waves.
In this paper, we will present complete N$^4$LO $NN$ potentials that also include the lower partial waves which receive contributions from contact interactions.

In Ref.~\cite{Ent15a}, we also demonstrated that the next-to-next-to-leading order (NNLO), 
the N$^3$LO, and the N$^4$LO contributions to the $NN$ interaction are all of about the same size, thus, not showing much of a trend towards convergence.
Therefore, in Ref.~\cite{Ent15b} we calculated the N$^5$LO (sixth order) contribution
which, indeed, turned out to be small.
The latter result may be perceived as an indication of convergence showing up at N$^5$LO.
This adds to the significance of order N$^4$LO.

Besides the above, we are faced with another set of convergence issues: The convergence
of the predictions for the properties of nuclear few- and many-body systems, in which also chiral
many-body forces are involved. To investigate these issues, one needs (besides those many-body forces) $NN$ potentials at all orders of chiral EFT, ranging from leading order (LO) to N$^4$LO,
and constructed consistently, i.~e., using the same power-counting scheme, consistent LECs, etc..

For that reason, we present in this paper $NN$ potentials through five orders from LO to
N$^4$LO, constructed with the above-stated consistencies and with a reproduction of the $NN$ data of the maximum quality possible at the respective orders. These potentials will allow 
for systematic investigations of nuclear few- and many-body systems with clear implications for
convergence and uncertainty quantifications (truncation errors)~\cite{FPW15,EKM15,Car16,MWF17}.

This paper is organized as follows:
In Sec.~II, 
we present the expansion of the $NN$ potential through all orders from
LO to N$^4$LO.
The reproduction 
of the $NN$ scattering data and the deuteron properties are given in Sec.~III. 
Some aspects regarding 3NFs are discussed in Sec.~IV, and uncertainty quantification is
considered in Sec.~V. Sec.~VI concludes the paper.

\section{Expansion of the $NN$ potential}
\label{sec_expans}

\subsection{Effective Langrangians}
\label{sec_lagr}

In the $\Delta$-less version of chiral EFT, which is the one we are pursuing here, the relevant degrees of freedom are
pions (Goldstone bosons) and nucleons.
Since the interactions of Goldstone bosons must
vanish at zero momentum transfer and in the chiral
limit ($m_\pi \rightarrow 0$), the low-energy expansion
of the effective Lagrangian is arranged in powers of derivatives
and pion masses.
This effective Lagrangian is subdivided into the following pieces,
\begin{equation}
{\cal L}_{\rm eff}
=
{\cal L}_{\pi\pi} 
+
{\cal L}_{\pi N} 
+
{\cal L}_{NN} 
 + \, \ldots \,,
\end{equation}
where ${\cal L}_{\pi\pi}$
deals with the dynamics among pions, 
${\cal L}_{\pi N}$ 
describes the interaction
between pions and a nucleon,
and ${\cal L}_{NN}$  contains two-nucleon contact interactions
which consist of four nucleon-fields (four nucleon legs) and no
meson fields.
The ellipsis stands for terms that involve two nucleons plus
pions and three or more
nucleons with or without pions, relevant for nuclear
many-body forces.
The individual Lagrangians are organized in terms of increasing orders:
\begin{eqnarray}
{\cal L}_{\pi\pi} 
 & = &
{\cal L}_{\pi\pi}^{(2)} 
+
{\cal L}_{\pi\pi}^{(4)}
 + \ldots \,, \\
{\cal L}_{\pi N} 
 & = &
{\cal L}_{\pi N}^{(1)} 
+
{\cal L}_{\pi N}^{(2)} 
+
{\cal L}_{\pi N}^{(3)} 
+
{\cal L}_{\pi N}^{(4)} 
+ \ldots , \\
\label{eq_LNN}
{\cal L}_{NN} &  = &
{\cal L}^{(0)}_{NN} +
{\cal L}^{(2)}_{NN} +
{\cal L}^{(4)}_{NN} + 
\ldots \,,
\end{eqnarray}
where the superscript refers to the number of derivatives or 
pion mass insertions (chiral dimension)
and the ellipses stand for terms of higher dimensions.
We use the heavy-baryon formulation of the Lagrangians, the
explicit expressions of which can be found in Refs.~\cite{ME11,KGE12}.

\subsection{Power counting}
\label{sec_pow}

Based upon the above Langrangians, an infinite number of diagrams contributing to the interactions among 
nucleons can be drawn. 
Nuclear potentials are defined by the irreducible types of these
graphs.
By definition, an irreducible graph is a diagram that
cannot be separated into two
by cutting only nucleon lines.
These graphs are then analyzed in terms of powers of small external momenta over the large scale: 
$(Q/\Lambda_\chi)^\nu$, where $Q$ is generic for a momentum (nucleon three-momentum
or pion four-momentum) or a pion mass and $\Lambda_\chi \sim 1$ GeV is the chiral symmetry breaking scale (hardronic scale, hard scale). Determining the power $\nu$ has become know
as power counting.

Following the Feynman rules of covariant perturbation theory,
a nucleon propagator is $Q^{-1}$,
a pion propagator $Q^{-2}$,
each derivative in any interaction is $Q$,
and each four-momentum integration $Q^4$.
This is also known as naive dimensional analysis or Weinberg counting.

Since we use the heavy-baryon formalism, we encounter terms which include factors of
$Q/M_N$, where $M_N$ denotes the nucleon mass.
We count the order of such terms by the rule
$Q/M_N \sim (Q/\Lambda_\chi)^2$,
for reasons explained in Ref.~\cite{Wei90}.

Applying some topological identities, one obtains
for the power of a connected irreducible diagram
involving $A$ nucleons~\cite{ME11,Wei90}
\begin{equation} \nu = -2 +2A  - 2C + 2L 
+ \sum_i \Delta_i \, ,
\label{eq_nu} 
\end{equation}
with
\begin{equation}
\Delta_i  \equiv   d_i + \frac{n_i}{2} - 2  \, ,
\label{eq_Deltai}
\end{equation}
where
$L$ denotes the number of loops in the diagram;
$d_i$ is the number of derivatives or pion-mass insertions 
and $n_i$ the number of nucleon fields (nucleon legs)
involved in vertex $i$;
the sum runs over all vertexes $i$ contained in the connected diagram 
under consideration.
Note that $\Delta_i \geq 0$
for all interactions allowed by chiral symmetry.

An important observation from power counting is that
the powers are bounded from below and, 
specifically, $\nu \geq 0$. 
This fact is crucial for the convergence of 
the low-momentum expansion.

Furthermore, the power formula 
Eq.~(\ref{eq_nu}) 
allows to predict
the leading orders of connected multi-nucleon forces.
Consider a $m$-nucleon irreducibly connected diagram
($m$-nucleon force) in an $A$-nucleon system ($m\leq A$).
The number of separately connected pieces is
$C=A-m+1$. Inserting this into
Eq.~(\ref{eq_nu}) together with $L=0$ and 
$\sum_i \Delta_i=0$ yields
$\nu=2m-4$. Thus, two-nucleon forces ($m=2$) appear
at $\nu=0$, three-nucleon forces ($m=3$) at
$\nu=2$ (but they happen to cancel at that order),
and four-nucleon forces at $\nu=4$ (they don't cancel).

For an irreducible 
$NN$ diagram ($A=2$, $C=1$), the
power formula collapses to the very simple expression
\begin{equation}
\nu =  2L + \sum_i \Delta_i \,.
\label{eq_nunn}
\end{equation}

In summary, the chief point of the ChPT expansion of the potential is
that,
at a given order $\nu$, there exists only a finite number
of graphs. This is what makes the theory calculable.
The expression $(Q/\Lambda_\chi)^{\nu+1}$ provides an estimate
of the relative size of the contributions left out and, thus,
of the relative uncertainty at order $\nu$.
The ability to calculate observables (in 
principle) to any degree of accuracy gives the theory 
its predictive power.

\begin{figure}[t]\centering
\scalebox{0.80}{\includegraphics{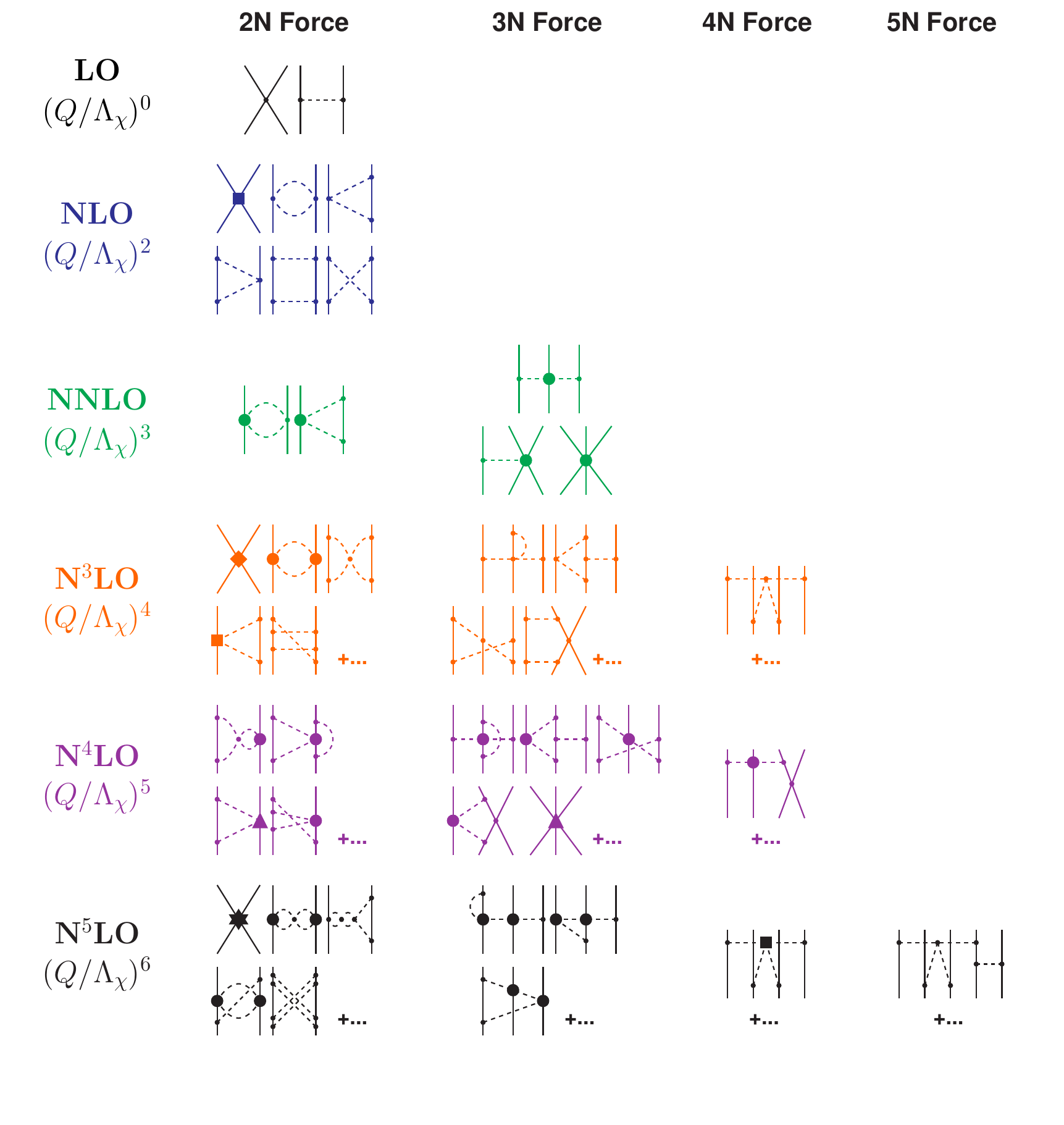}}
\vspace*{-0.5cm}
\caption{Hierarchy of nuclear forces in ChPT. Solid lines
represent nucleons and dashed lines pions. 
Small dots, large solid dots, solid squares, triangles, diamonds, and stars
denote vertexes of index $\Delta_i= \, $ 0, 1, 2, 3, 4, and 6, respectively. 
Further explanations are
given in the text.}
\label{fig_hi}
\end{figure}

Chiral perturbation theory and power counting
imply that nuclear forces evolve as a hierarchy
controlled by the power $\nu$, see Fig.~\ref{fig_hi} for an overview.
In what follows, we will focus on the two-nucleon force (2NF).

\subsection{The long-range $NN$ potential}
\label{sec_long}

The long-range part of the $NN$ potential is built up from pion exchanges,
which are ruled by chiral symmetry.
The various pion-exchange contributions may be analyzed
according to the number of pions being exchanged between the two
nucleons:
\begin{equation}
V = V_{1\pi} + V_{2\pi} + V_{3\pi} + \ldots \,,
\end{equation}
where the meaning of the subscripts is obvious
and the ellipsis represents $4\pi$ and higher pion exchanges. For each of the above terms, 
we have a low-momentum expansion:
\begin{eqnarray}
V_{1\pi} & = & V_{1\pi}^{(0)} + V_{1\pi}^{(2)} 
+ V_{1\pi}^{(3)} + V_{1\pi}^{(4)} + V_{1\pi}^{(5)} + \ldots 
\label{eq_1pe_orders}
\\
V_{2\pi} & = & V_{2\pi}^{(2)} + V_{2\pi}^{(3)} + V_{2\pi}^{(4)} + V_{2\pi}^{(5)} 
+ \ldots \\
V_{3\pi} & = & V_{3\pi}^{(4)} + V_{3\pi}^{(5)} + \ldots \,,
\end{eqnarray}
where the superscript denotes the order $\nu$ of the expansion.

Order by order, the long-range $NN$ potential builds up as follows:
\beqa
V_{\rm LO} & \equiv & V^{(0)} =
V_{1\pi}^{(0)} 
\label{eq_VLO}
\\
V_{\rm NLO} & \equiv & V^{(2)} = V_{\rm LO} +
V_{1\pi}^{(2)} +
V_{2\pi}^{(2)} 
\label{eq_VNLO}
\\
V_{\rm NNLO} & \equiv & V^{(3)} = V_{\rm NLO} +
V_{1\pi}^{(3)} + 
V_{2\pi}^{(3)} 
\label{eq_VNNLO}
\\
V_{\rm N3LO} & \equiv & V^{(4)} = V_{\rm NNLO} +
V_{1\pi}^{(4)} + 
V_{2\pi}^{(4)} +
V_{3\pi}^{(4)} 
\label{eq_VN3LO}
\\
V_{\rm N4LO} & \equiv & V^{(5)} = V_{\rm N3LO} +
V_{1\pi}^{(5)} + 
V_{2\pi}^{(5)} +
V_{3\pi}^{(5)} 
\label{eq_VN4LO}
\eeqa
where 
LO stands for leading order, NLO for next-to-leading order, etc..

\subsubsection{Leading order}

At leading order, only one-pion exchange (1PE) contributes to the long range, cf.\ Fig.~\ref{fig_hi}.
The charge-independent 1PE is given by
\begin{equation}
V_{1\pi}^{\rm(CI)} ({\vec p}~', \vec p) = - 
\frac{g_A^2}{4f_\pi^2}
\: 
\bm{\tau}_1 \cdot \bm{\tau}_2 
\:
\frac{
\vec \sigma_1 \cdot \vec q \,\, \vec \sigma_2 \cdot \vec q}
{q^2 + m_\pi^2} 
\,,
\label{eq_1PEci}
\end{equation}
where ${\vec p}\,'$ and $\vec p$ denote the final and initial nucleon momenta in the 
center-of-mass system, 
respectively. Moreover, $\vec q = {\vec p}\,' - \vec p$ is the momentum transfer, 
 and $\vec \sigma_{1,2}$ and $\bm{\tau}_{1,2}$ are the spin 
and isospin operators of nucleon 1 and 2, respectively. Parameters
$g_A$, $f_\pi$, and $m_\pi$ denote the axial-vector coupling constant,
pion-decay constant, and the pion mass, respectively. See Table~\ref{tab_basic}
for their values.  
Higher order corrections to the 1PE  are taken care of by  mass
and coupling constant renormalizations. Note also that, on 
shell, there are no relativistic corrections. Thus, we apply  1PE in the form
\eq{eq_1PEci} through all orders.

For the $NN$ potentials constructed in this paper, we take the charge-dependence of the 1PE due to pion-mass splitting into account. Thus, in proton-proton ($pp$) and neutron-neutron ($nn$) scattering, we actually use
\begin{equation}
V_{1\pi}^{(pp)} ({\vec p}~', \vec p) =
V_{1\pi}^{(nn)} ({\vec p}~', \vec p) 
= V_{1\pi} (m_{\pi^0})
\,,
\label{eq_1pepp}
\end{equation}
and in neutron-proton ($np$) scattering,
we apply
\begin{equation}
V_{1\pi}^{(np)} ({\vec p}~', \vec p) 
= -V_{1\pi} (m_{\pi^0}) + (-1)^{I+1}\, 2\, V_{1\pi} (m_{\pi^\pm})
\,,
\label{eq_1penp}
\end{equation}
where $I=0,1$ denotes the total isospin of the two-nucleon system and
\begin{equation}
V_{1\pi} (m_\pi) \equiv - \,
\frac{g_A^2}{4f_\pi^2} \,
\frac{
\vec \sigma_1 \cdot \vec q \,\, \vec \sigma_2 \cdot \vec q}
{q^2 + m_\pi^2} 
\,.
\end{equation}
Formally speaking, the charge-dependence of the 1PE exchange is of order 
NLO~\cite{ME11}, but we include it also at leading order to make the comparison with
the (charge-dependent) phase-shift analyses meaningful.

\begin{table}[t]
\caption{Basic constants used throughout this work~\cite{PDG}.}
\label{tab_basic}
\smallskip
\begin{tabular}{lcl}
\hline 
\hline 
\noalign{\smallskip}
  Quantity            &  \hspace{2cm} & Value \\
\hline
\noalign{\smallskip}
Axial-vector coupling constant $g_A$ && 1.29 \\
Pion-decay constant $f_\pi$ && 92.4 MeV \\
Charged-pion mass $m_{\pi^\pm}$ && 139.5702 MeV \\
Neutral-pion mass $m_{\pi^0}$ && 134.9766 MeV \\
Average pion-mass $\bar{m}_\pi$ && 138.0390 MeV \\
Proton mass $M_p$ && 938.2720 MeV \\
Neutron mass $M_n$ && 939.5654 MeV \\
Average nucleon-mass $\bar{M}_N$ && 938.9183 MeV \\
\hline
\hline
\noalign{\smallskip}
\end{tabular}
\end{table}

\subsubsection{Subleading pion exchanges}

Two-pion exchange starts at NLO and continues through all higher orders.
In Fig.~\ref{fig_hi}, the corresponding diagrams are show completely up to NNLO.
Beyond that order, the number of diagrams increases so dramatically that 
we show only a few symbolic graphs. The situation is similar for the 3PE contributions which
start at N$^3$LO. Also the mathematical formulas are getting increasingly involved.
A complete collection of all formulas concerning the
2PE and 3PE contributions through all orders from 
NLO to N$^4$LO is given in Ref.~\cite{Ent15a}. Therefore, we will not reprint 
the complicated math here and refer the interested reader to the comprehensive compendium~\cite{Ent15a}.
In all 2PE and 3PE contributions, we use the average pion mass,
$\bar{m}_\pi=138.039$ MeV. The charge-dependence caused by pion-mass splitting in 2PE has been
found to be negligible in all partial waves with $L>0$~\cite{LM98b}. The small effect in $^1S_0$ is absorbed into the charge-dependence of the zeroth-order contact parameter
$\widetilde{C}_{^1 S_0}$, see below.

The contributions have the following general decomposition:
\begin{eqnarray} 
V({\vec p}~', \vec p) &  = &
 \:\, V_C \:\, + \bm{\tau}_1 \cdot \bm{\tau}_2 \, W_C 
\nonumber \\ & + &
\left[ \, V_S \:\, + \bm{\tau}_1 \cdot \bm{\tau}_2 \, W_S 
\,\:\, \right] \,
\vec\sigma_1 \cdot \vec \sigma_2
\nonumber \\ &+& 
\left[ \, V_{LS} + \bm{\tau}_1 \cdot \bm{\tau}_2 \, W_{LS}    
\right] \,
\left(-i \vec S \cdot (\vec q \times \vec k) \,\right)
\nonumber \\ &+& 
\left[ \, V_T \:\,     + \bm{\tau}_1 \cdot \bm{\tau}_2 \, W_T 
\,\:\, \right] \,
\vec \sigma_1 \cdot \vec q \,\, \vec \sigma_2 \cdot \vec q  
\, ,
\label{eq_nnamp}
\end{eqnarray}
where $\vec k =({\vec p}\,' + \vec p)/2$ denotes the average momentum and $\vec S =(\vec\sigma_1+\vec\sigma_2)/2 $ is the total spin.
For on-shell scattering, $V_\alpha$ and $W_\alpha$ ($\alpha=C,S,LS,T$) can be 
expressed as functions of $q= |\vec q\,|$. 
  
We consider
loop contributions in terms of their spectral functions, from which
the momentum-space amplitudes $V_\alpha(q)$ and $W_\alpha(q)$
are obtained
via the subtracted dispersion integrals:
\begin{eqnarray} 
V_{C,S}(q) &=& 
-{2 q^6 \over \pi} \int_{nm_\pi}^{\tilde{\Lambda}} d\mu \,
{{\rm Im\,}V_{C,S}(i \mu) \over \mu^5 (\mu^2+q^2) }\,, 
\nn
V_{T,LS}(q) &=& 
{2 q^4 \over \pi} \int_{nm_\pi}^{\tilde{\Lambda}} d\mu \,
{{\rm Im\,}V_{T,LS}(i \mu) \over \mu^3 (\mu^2+q^2) }\,, 
\label{eq_disp}
\end{eqnarray}
and similarly for $W_{C,S,T,LS}$.
The thresholds are given by
$n=2$ for two-pion exchange and $n=3$ for
three-pion exchange.
For $\tilde{\Lambda} \rightarrow \infty$ the above dispersion integrals yield the
results of dimensional regularization, while for finite $\tilde{\Lambda} \geq nm_\pi$
we employ the method known as spectral-function regularization (SFR) \cite{EGM04}. The 
purpose of the finite scale $\tilde{\Lambda}$ is to constrain the imaginary parts to the  
low-momentum region where chiral effective field theory is applicable.  
Thus, a reasonable choice for $\tilde{\Lambda}$ is to keep it below the masses of the vector mesons
$\rho(770)$ and $\omega(782)$, but above the $f_0(500)$ [also know as $\sigma(500)$]~\cite{PDG}.
This suggests that the region 600-700 MeV is appropriate for $\tilde{\Lambda}$.
Consequently, we use $\tilde{\Lambda} =650$ MeV in all orders, except for N$^4$LO where we apply 700 MeV. We use this slightly larger value for N$^4$LO, because it is suggestive that higher orders may permit for an extension to higher momenta.

\subsubsection{The pion-nucleon low-energy constants}
\label{sec_LECs}

\begin{table}
\caption{The $\pi N$ LECs as determined in
the Roy-Steiner-equation analysis of $\pi N$ scattering conducted in  Ref.~\cite{Hof15}.
The given orders of the chiral expansion refer to the $NN$ system. Note that the orders, 
 at which the LECs are extracted from the $\pi N$ system, are always lower by one order as compared of the $NN$ system in which the LECs are applied.
The $c_i$, $\bar{d}_i$, 
and $\bar{e}_i$ are the LECs of the second, third, and fourth order $\pi N$ Lagrangian~\cite{KGE12} and are 
 in units of GeV$^{-1}$, GeV$^{-2}$, and GeV$^{-3}$, respectively.
The uncertainties in the last digits are given in parentheses after the values.}
\label{tab_lecs}
\smallskip
\begin{tabular*}{\textwidth}{@{\extracolsep{\fill}}crrr}
\hline 
\hline 
\noalign{\smallskip}
              & NNLO & N$^3$LO & N$^4$LO \\
\hline
\noalign{\smallskip}
$c_1$ & --0.74(2) & --1.07(2) & --1.10(3) \\
$c_2$ & ---  & 3.20(3) & 3.57(4) \\
$c_3$ & --3.61(5) & --5.32(5) & --5.54(6) \\
$c_4$ & 2.44(3) & 3.56(3) & 4.17(4) \\
$\bar{d}_1 + \bar{d}_2$ & --- & 1.04(6) & 6.18(8) \\
$\bar{d}_3$ & --- & --0.48(2) & --8.91(9) \\
$\bar{d}_5$ & --- & 0.14(5) & 0.86(5) \\
$\bar{d}_{14} - \bar{d}_{15}$ & --- & --1.90(6) & --12.18(12) \\
$\bar{e}_{14}$ & --- & --- & 1.18(4) \\
$\bar{e}_{17}$ & --- & --- & --0.18(6) \\
\hline
\hline
\noalign{\smallskip}
\end{tabular*}
\end{table}

Chiral symmetry establishes a link between the dynamics in the $\pi N$-system 
and the $N\!N$-system through common low-energy constants. 
Therefore, consistency requires that we use the LECs for subleading $\pi N$-couplings as 
determined in analysis of low-energy $\pi N$-scattering.
Over the years, there have been many such determinations of questionable reliability.
Fortunately, that has changed recently with the analysis 
by Hoferichter and Ruiz de Elvira~\cite{note1}
{\it et al.}~\cite{Hof15}, in
which the Roy-Steiner (RS) equations are applied.
The RS equations are a set of coupled partial-wave dispersion relations constraint by analyticity, unitarity, and crossing symmetry. In the work of Ref.~\cite{Hof15}, they are used
to extract the LECs from the subthreshold point in
$\pi N$ scattering instead of the physical region. This is the preferred method for LECs
to be applied in chiral potentials where, e.~g., a one-loop $\pi N$ amplitude
 leads to a two-loop contribution in $NN$. 
 Such diagrams are best evaluated by means of Cutkosky rules~\cite{Kai01a,Ent15a,Ent15b}.
 The $\pi N$ amplitude that enters the dispersion integrals is weighted much closer to 
 subthreshold kinematics than to the threshold point.
The LECs determined in Ref.~\cite{Hof15}
carry very small uncertainties 
(cf.\ Table~\ref{tab_lecs}) for, essentially, two reasons: first, because of the constraints built into the RS equations; second, because of the use of the high-accuracy $\pi N$ scattering lengths extracted from pionic atoms.
In fact, the uncertainties are so small that they are negligible for our purposes.
This makes the variation of the $\pi N$ LECs in $NN$ potential construction obsolete
and reduces the error budget in applications of these potentials. 
For the potentials constructed in this paper, the central values of Table~\ref{tab_lecs} are applied.

\subsection{The short-range $NN$ potential}
\label{sec_short}

The short-range $NN$ potential is described by contributions of the contact type,
which are constrained by parity, time-reversal, and the usual invariances, but not by chiral symmetry. Terms that include a factor
$\bm{\tau}_1 \cdot \bm{\tau}_2$ (owing to isospin invariance) can be left out due to Fierz ambiguity. 
Because of  parity and time-reversal only even powers of momentum
are allowed.
Thus, the expansion of the contact potential is
formally written as
\begin{equation}
V_{\rm ct} =
V_{\rm ct}^{(0)} + 
V_{\rm ct}^{(2)} + 
V_{\rm ct}^{(4)} + 
V_{\rm ct}^{(6)} 
+ \ldots \; ,
\label{eq_ct}
\end{equation}
where the superscript denotes the power or order.

The zeroth order (leading order, LO) contact potential is given by
\begin{equation}
V_{\rm ct}^{(0)}(\vec{p'},\vec{p}) =
C_S +
C_T \, \vec{\sigma}_1 \cdot \vec{\sigma}_2 \, 
\label{eq_ct0}
\end{equation}
and, in terms of partial waves, 
\be
V_{\rm ct}^{(0)}(^1 S_0)          &=&  \widetilde{C}_{^1 S_0} =
4\pi\, ( C_S - 3 \, C_T )
\label{eq_ct0_1s0}
\\
V_{\rm ct}^{(0)}(^3 S_1)          &=&  \widetilde{C}_{^3 S_1} =
4\pi\, ( C_S + C_T ) \,.
\label{eq_ct0_3s1}
\ee
To deal with the isospin breaking in the $^1S_0$ state, we treat $\widetilde{C}_{^1 S_0}$
in a charge-dependent way.
Thus, we will distinguish between $\widetilde{C}_{^1 S_0}^{\rm pp}$,
$\widetilde{C}_{^1 S_0}^{\rm np}$, and $\widetilde{C}_{^1 S_0}^{\rm nn}$.

At second order (NLO), we have
\be
V_{\rm ct}^{(2)}(\vec{p'},\vec{p}) &=&
C_1 \, q^2 +
C_2 \, k^2 
\nonumber 
\\ &+& 
\left(
C_3 \, q^2 +
C_4 \, k^2 
\right) \vec{\sigma}_1 \cdot \vec{\sigma}_2 
\nonumber 
\\
&+& C_5 \left( -i \vec{S} \cdot (\vec{q} \times \vec{k}) \right)
\nonumber 
\\ &+& 
 C_6 \, ( \vec{\sigma}_1 \cdot \vec{q} )\,( \vec{\sigma}_2 \cdot 
\vec{q} )
\nonumber 
\\ &+& 
 C_7 \, ( \vec{\sigma}_1 \cdot \vec{k} )\,( \vec{\sigma}_2 \cdot 
\vec{k} ) \,,
\label{eq_ct2}
\ee
and partial-wave decomposition yields
\be
V_{\rm ct}^{(2)}(^1 S_0)          &=&  C_{^1 S_0} ( p^2 + {p'}^2 ) 
\nonumber \\
V_{\rm ct}^{(2)}(^3 P_0)          &=&  C_{^3 P_0} \, p p'
\nonumber \\
V_{\rm ct}^{(2)}(^1 P_1)          &=&  C_{^1 P_1} \, p p' 
\nonumber \\
V_{\rm ct}^{(2)}(^3 P_1)          &=&  C_{^3 P_1} \, p p' 
\nonumber \\
V_{\rm ct}^{(2)}(^3 S_1)          &=&  C_{^3 S_1} ( p^2 + {p'}^2 ) 
\nonumber \\
V_{\rm ct}^{(2)}(^3 S_1- ^3 D_1)  &=&  C_{^3 S_1- ^3 D_1}  p^2 
\nonumber \\
V_{\rm ct}^{(2)}(^3 D_1- ^3 S_1)  &=&  C_{^3 S_1- ^3 D_1}  {p'}^2 
\nonumber \\
V_{\rm ct}^{(2)}(^3 P_2)          &=&  C_{^3 P_2} \, p p' 
\,.
\label{eq_ct2_pw}
\ee
The relationship between the $C_{^{(2S+1)}L_J}$ and the $C_i$ can be found in Ref.~\cite{ME11}.

The fourth order (N$^3$LO) contacts are
\be
V_{\rm ct}^{(4)}(\vec{p'},\vec{p}) &=&
D_1 \, q^4 +
D_2 \, k^4 +
D_3 \, q^2 k^2 +
D_4 \, (\vec{q} \times \vec{k})^2 
\nonumber 
\\ &+& 
\left(
D_5 \, q^4 +
D_6 \, k^4 +
D_7 \, q^2 k^2 +
D_8 \, (\vec{q} \times \vec{k})^2 
\right) \vec{\sigma}_1 \cdot \vec{\sigma}_2 
\nonumber 
\\ &+& 
\left(
D_9 \, q^2 +
D_{10} \, k^2 
\right) \left( -i \vec{S} \cdot (\vec{q} \times \vec{k}) \right)
\nonumber 
\\ &+& 
\left(
D_{11} \, q^2 +
D_{12} \, k^2 
\right) ( \vec{\sigma}_1 \cdot \vec{q} )\,( \vec{\sigma}_2 
\cdot \vec{q})
\nonumber 
\\ &+& 
\left(
D_{13} \, q^2 +
D_{14} \, k^2 
\right) ( \vec{\sigma}_1 \cdot \vec{k} )\,( \vec{\sigma}_2 
\cdot \vec{k})
\nonumber 
\\ &+& 
D_{15} \left( 
\vec{\sigma}_1 \cdot (\vec{q} \times \vec{k}) \, \,
\vec{\sigma}_2 \cdot (\vec{q} \times \vec{k}) 
\right) 
\,,
\label{eq_ct4}
\ee
with contributions by partial waves,
\be
V_{\rm ct}^{(4)}(^1 S_0)          &=&  \widehat{D}_{^1 S_0}          
({p'}^4 + p^4) +
                              D_{^1 S_0}          {p'}^2 p^2 
\nonumber 
\\
V_{\rm ct}^{(4)}(^3 P_0)          &=&        D_{^3 P_0}          
({p'}^3 p + p' p^3) 
\nonumber 
\\
V_{\rm ct}^{(4)}(^1 P_1)          &=&        D_{^1 P_1}          
({p'}^3 p + p' p^3) 
\nonumber 
\\
V_{\rm ct}^{(4)}(^3 P_1)          &=&        D_{^3 P_1}          
({p'}^3 p + p' p^3) 
\nonumber 
\\
V_{\rm ct}^{(4)}(^3 S_1)          &=&  \widehat{D}_{^3 S_1}          
({p'}^4 + p^4) +
                              D_{^3 S_1}          {p'}^2 p^2 
\nonumber 
\\
V_{\rm ct}^{(4)}(^3 D_1)          &=&        D_{^3 D_1}          
{p'}^2 p^2 
\nonumber 
\\
V_{\rm ct}^{(4)}(^3 S_1 - ^3 D_1) &=&  \widehat{D}_{^3 S_1 - ^3 D_1} 
p^4             +
                              D_{^3 S_1 - ^3 D_1} {p'}^2 p^2
\nonumber 
\\
V_{\rm ct}^{(4)}(^3 D_1 - ^3 S_1) &=&  \widehat{D}_{^3 S_1 - ^3 D_1} 
{p'}^4             +
                              D_{^3 S_1 - ^3 D_1} {p'}^2 p^2
\nonumber 
\\
V_{\rm ct}^{(4)}(^1 D_2)          &=&        D_{^1 D_2}          
{p'}^2 p^2 
\nonumber 
\\
V_{\rm ct}^{(4)}(^3 D_2)          &=&        D_{^3 D_2}          
{p'}^2 p^2 
\nonumber 
\\
V_{\rm ct}^{(4)}(^3 P_2)          &=&        D_{^3 P_2}          
({p'}^3 p + p' p^3) 
\nonumber 
\\
V_{\rm ct}^{(4)}(^3 P_2 - ^3 F_2) &=&        D_{^3 P_2 - ^3 F_2} {p'}p^3
\nonumber 
\\
V_{\rm ct}^{(4)}(^3 F_2 - ^3 P_2) &=&        D_{^3 P_2 - ^3 F_2} {p'}^3p
\nonumber 
\\
V_{\rm ct}^{(4)}(^3 D_3)          &=&        D_{^3 D_3}          
{p'}^2 p^2 
\,.
\label{eq_ct4_pw}
\ee
Reference~\cite{ME11} provides formulas that relate 
the $D_{^{(2S+1)}L_J}$ to the $D_i$. 

The next higher order is sixth order (N$^5$LO) at which, finally, also $F$-waves
are affected in the following way:
\be
V_{\rm ct}^{(6)}(^3 F_2)          &=&        E_{^3 F_2}    {p'}^3 p^3 
\nonumber 
\\
V_{\rm ct}^{(6)}(^1 F_3)          &=&        E_{^1 F_3}    {p'}^3 p^3 
\nonumber 
\\
V_{\rm ct}^{(6)}(^3 F_3)          &=&        E_{^3 F_3}    {p'}^3 p^3 
\nonumber 
\\
V_{\rm ct}^{(6)}(^3 F_4)          &=&        E_{^3 F_4}    {p'}^3 p^3 
\,.
\label{eq_ct6}
\ee
To obtain an optimal fit of the $NN$ data at the highest order we consider in this paper, 
we include the above $F$-wave contacts in our N$^4$LO potentials.

\subsection{Charge dependence}
\label{sec_CD}

This is to summarize what charge-dependence we include.
Through all orders, we take the charge-dependence of the 1PE due to pion-mass splitting
into account, Eqs.~(\ref{eq_1pepp}) and (\ref{eq_1penp}).
Charge-dependence is seen most prominently in the $^1S_0$ state at low energies, particularly, in the $^1S_0$ scattering lengths. Charge-dependent 1PE cannot explain it all. 
The remainder is accounted for by treating the $^1S_0$ LO contact parameter, $\widetilde{C}_{^1 S_0}$, Eq.~(\ref{eq_ct0_1s0}),
in a charge-dependent way. 
Thus, we will distinguish between $\widetilde{C}_{^1 S_0}^{\rm pp}$,
$\widetilde{C}_{^1 S_0}^{\rm np}$, and $\widetilde{C}_{^1 S_0}^{\rm nn}$.
For $pp$ scattering at any order, we include the relativistic Coulomb potential~\cite{AS83,Ber88}.
Finally, at N$^3$LO and N$^4$LO, we take into account irreducible $\pi$-$\gamma$ 
exchange~\cite{Kol98},
which affects only the $np$ potential.
We also take nucleon-mass splitting into account,
or in other words,
we always apply the correct values for the masses of the nucleons involved in the various charge-dependent $NN$ potentials.

For a comprehensive discussion of all possible sources for the charge-dependence of the $NN$
interaction, see Ref.~\cite{ME11}.

\subsection{The full potential}
\label{sec_pot}

The sum of long-range [Eqs.~(\ref{eq_VLO})-(\ref{eq_VN4LO})]
 plus short-range potentials [Eq.~(\ref{eq_ct})]  results in:
\beqa
V_{\rm LO} & \equiv & V^{(0)} =
V_{1\pi} + V_{\rm ct}^{(0)} 
\label{eq_VVLO}
\\
V_{\rm NLO} & \equiv & V^{(2)} = V_{\rm LO} 
+V_{2\pi}^{(2)} 
+ V_{\rm ct}^{(2)} 
\label{eq_VVNLO}
\\
V_{\rm NNLO} & \equiv & V^{(3)} = V_{\rm NLO} 
+V_{2\pi}^{(3)} 
\label{eq_VVNNLO}
\\
V_{\rm N3LO} & \equiv & V^{(4)} = V_{\rm NNLO} +
V_{2\pi}^{(4)} +
V_{3\pi}^{(4)} 
 + V_{\rm ct}^{(4)} 
\label{eq_VVN3LO}
\\
V_{\rm N4LO} & \equiv & V^{(5)} = V_{\rm N3LO} + 
V_{2\pi}^{(5)} +
V_{3\pi}^{(5)} 
\,,
\label{eq_VVN4LO}
\eeqa
where we left out the higher order corrections to the 1PE because, as discussed, 
they are absorbed by mass and coupling constant renormalizations.
It is also understood that the charge-dependence discussed in the previous subsection is 
included.

In our systematic potential construction, we follow the above scheme, except for 
two physically motivated modifications.
We add to $V_{\rm N3LO}$ the $1/M_N$ correction of the NNLO 2PE proportional to $c_i$.
This correction is proportional to $c_i/M_N$ and appears nominally at fifth order,
because we count $Q/M_N \sim (Q/\Lambda_\chi)^2$. This contribution
 is given in Eqs.\ (2.19)-(2.23) of Ref.~\cite{Ent15a} and we denote it by $V_{2\pi,c_i/M_N}^{(5)}$.
In short, in Eq.~ (\ref{eq_VVN3LO}), we replace
\begin{equation}
V_{\rm N3LO} 
\longmapsto
V_{\rm N3LO} + V_{2\pi,c_i/M_N}^{(5)}
\,.
\end{equation}
As demonstrated in Ref.~\cite{EM02}, the 2PE bubble diagram proportional to $c_i^2$
that appears at N$^3$LO is unrealistically attractive, while the $c_i/M_N$ correction is large
and repulsive. Therefore, it makes sense to group these diagrams together to arrive at a more
realistic intermediate attraction at N$^3$LO.

The second modification consists of adding to $V_{\rm N4LO}$ the four $F$-wave contacts
listed in Eq.~(\ref{eq_ct6}) to ensure an optimal fit of the $NN$ data for the potential of the highest order constructed in this work.

The potential $V$ is, in principal, an invariant amplitude (with relativity taken into account perturbatively) and, thus, satisfies a relativistic scattering equation, like, e.\ g., the
Blankenbeclar-Sugar (BbS) equation~\cite{BS66},
which reads explicitly,
\begin{equation}
{T}({\vec p}~',{\vec p})= {V}({\vec p}~',{\vec p})+
\int \frac{d^3p''}{(2\pi)^3} \:
{V}({\vec p}~',{\vec p}~'') \:
\frac{M_N^2}{E_{p''}} \:  
\frac{1}
{{ p}^{2}-{p''}^{2}+i\epsilon} \:
{T}({\vec p}~'',{\vec p}) 
\label{eq_bbs2}
\end{equation}
with $E_{p''}\equiv \sqrt{M_N^2 + {p''}^2}$ and $M_N$ the nucleon mass.
The advantage of using a relativistic scattering equation is that it automatically
includes relativistic kinematical corrections to all orders. Thus, in the scattering equation,
no propagator modifications are necessary when moving up to higher orders.

Defining
\begin{equation}
\widehat{V}({\vec p}~',{\vec p})
\equiv 
\frac{1}{(2\pi)^3}
\sqrt{\frac{M_N}{E_{p'}}}\:  
{V}({\vec p}~',{\vec p})\:
 \sqrt{\frac{M_N}{E_{p}}}
\label{eq_minrel1}
\end{equation}
and
\begin{equation}
\widehat{T}({\vec p}~',{\vec p})
\equiv 
\frac{1}{(2\pi)^3}
\sqrt{\frac{M_N}{E_{p'}}}\:  
{T}({\vec p}~',{\vec p})\:
 \sqrt{\frac{M_N}{E_{p}}}
\,,
\label{eq_minrel2}
\end{equation}
where the factor $1/(2\pi)^3$ is added for convenience,
the BbS equation collapses into the usual, nonrelativistic
Lippmann-Schwinger (LS) equation,
\begin{equation}
 \widehat{T}({\vec p}~',{\vec p})= \widehat{V}({\vec p}~',{\vec p})+
\int d^3p''\:
\widehat{V}({\vec p}~',{\vec p}~'')\:
\frac{M_N}
{{ p}^{2}-{p''}^{2}+i\epsilon}\:
\widehat{T}({\vec p}~'',{\vec p}) \, .
\label{eq_LS}
\end{equation}
Since 
$\widehat V$ 
satisfies Eq.~(\ref{eq_LS}), 
it may be regarded as a nonrelativistic potential. By the same token, 
$\widehat{T}$ 
may be considered as the nonrelativistic 
T-matrix.
All technical aspects associated with the solution of the LS equation 
can be found in Appendix A of Ref.~\cite{Mac01}, including 
specific formulas for the calculation of the $np$ and $pp$ phase shifts (with Coulomb).
Additional details 
concerning the relevant operators and their decompositions 
are given in section~4 of Ref.~\cite{EAH71}. Finally, computational methods
to solve the LS equation are found in Ref.~\cite{Mac93}.

\subsection{Regularization and non-perturbative renormalization}
\label{sec_reno}

Iteration of $\widehat V$ in the LS equation, Eq.~(\ref{eq_LS}),
requires cutting $\widehat V$ off for high momenta to avoid infinities.
This is consistent with the fact that ChPT
is a low-momentum expansion which
is valid only for momenta $Q < \Lambda_\chi \approx 1$ GeV.
Therefore, the potential $\widehat V$
is multiplied
with the regulator function $f(p',p)$,
\begin{equation}
{\widehat V}(\vec{ p}~',{\vec p})
\longmapsto
{\widehat V}(\vec{ p}~',{\vec p}) \, f(p',p) 
\end{equation}
with
\begin{equation}
f(p',p) = \exp[-(p'/\Lambda)^{2n}-(p/\Lambda)^{2n}] \,,
\label{eq_f}
\end{equation}
such that
\begin{equation}
{\widehat V}(\vec{ p}~',{\vec p}) \, f(p',p) 
\approx
{\widehat V}(\vec{ p}~',{\vec p})
\left\{1-\left[\left(\frac{p'}{\Lambda}\right)^{2n}
+\left(\frac{p}{\Lambda}\right)^{2n}\right]+ \ldots \right\} 
\,.
\label{eq_reg_exp}
\end{equation}
For the cutoff parameter $\Lambda$, we apply three different values, namely,
450, 500, and 550 MeV.

Equation~(\ref{eq_reg_exp}) provides an indication of the fact that
the exponential cutoff does not necessarily
affect the given order at which 
the calculation is conducted.
For sufficiently large $n$, the regulator introduces contributions that 
are beyond the given order. Assuming a good rate
of convergence of the chiral expansion, such orders are small 
as compared to the given order and, thus, do not
affect the accuracy at the given order.
Thus, we use $n=2$ for 3PE and 2PE and $n=4$ for 1PE (except in LO and NLO, where we use $n=2$ for 1PE).
For contacts of order $\nu$, $n$ is chosen such that $2n > \nu$.

In our calculations, we apply, of course,
the exponential form, Eq.~(\ref{eq_f}),
and not the expansion Eq.~(\ref{eq_reg_exp}). On a similar note, we also
do not expand the square-root factors
in Eqs.~(\ref{eq_minrel1}-\ref{eq_minrel2})
because they are kinematical factors which guarantee
relativistic elastic unitarity.

It is pretty obvious that results for the $T$-matrix may
depend sensitively on the regulator and its cutoff parameter.
The removal of such regulator dependence is known as renormalization.

The renormalization of the {\it perturbatively} calculated $NN$ potential
 is not a problem. {\it The problem
is nonperturbative renormalization.}
This problem typically occurs in {\it nuclear} EFT because
nuclear physics is characterized by bound states and large scattering length which
are nonperturbative in nature. Or in other words, to obtain the nuclear amplitude, the potential has to be resummed (to infinite orders) in the LS equation Eq.~(\ref{eq_LS}).
EFT power counting may be different for nonperturbative processes as
compared to perturbative ones. Such difference may be caused by the infrared
enhancement of the reducible diagrams generated in the LS equation.

Weinberg's implicit assumption~\cite{Wei90} was that the counterterms
introduced to renormalize the perturbatively calculated
potential, based upon naive dimensional analysis (``Weinberg counting'', cf.~Sec.~\ref{sec_pow}),
are also sufficient to renormalize the nonperturbative
resummation of the potential in the LS equation.

Weinberg's assumption may not be correct as first pointed out by Kaplan, Savage, and Wise~\cite{KSW96}, and we like to refer the interested reader to 
Section 4.5 of Ref.~\cite{ME11} for a comprehensive discussion of the issue.
Even today, no generally accepted solution to this problem has emerged and some more recent proposals can be found in Refs.~\cite{Bir06,LY12,Lon16,Val16,Val16a,San17,EGM17,Kon17}.
Concerning the construction of quantitative $NN$ potential
(by which we mean $NN$ potentials suitable for use in contemporary many-body nuclear methods), 
only Weinberg counting
has been used with success during the past 25 years~\cite{ORK94,EGM05,ME11,EKM15,Pia15,Eks15}, which is why also in the present work
we will apply Weinberg counting.

In spite of the criticism, Weinberg counting may be perceived as not unreasonable by the following argument.
For a successful EFT (in its domain of validity), one must be able to claim independence of the predictions on the regulator within the theoretical error.
Also,                                         
truncation errors must decrease as we go to higher and higher orders.
These are precisely the goals of renormalization.  

Lepage~\cite{Lep97} has stressed that the cutoff independence should be examined
for cutoffs below the hard scale and not beyond. Ranges of cutoff independence within the
theoretical error are to be identified using Lepage plots~\cite{Lep97}.
A systematic investigation of this kind has been conducted in Ref.~\cite{Mar13}.
In that work, the error of the predictions was quantified by calculating the $\chi^2$/datum 
for the reproduction of the $np$ elastic scattering data
as a function of the cutoff parameter $\Lambda$ of the regulator function
Eq.~(\ref{eq_f}). Predictions by chiral $np$ potentials at 
order NLO and NNLO were investigated applying Weinberg counting 
for the counter terms ($NN$ contact terms).
It is found that the reproduction of the $np$ data at lab.\ energies below 200 MeV is generally poor
at NLO, while at NNLO the $\chi^2$/datum assumes acceptable values (a clear demonstration of
order-by-order improvement). Moreover, at NNLO, 
a ``plateau'' of constant low $\chi^2$ for
cutoff parameters ranging from about 450 to 850 MeV can be identified. This may be perceived as cutoff independence
(and, thus, successful renormalization) for the relevant range of cutoff parameters.

\begin{table}
\caption{Publication history of the $NN$ data below 350 MeV laboratory energy
and references for their listings.
Only data that pass the Nijmegen acceptance criteria~\cite{Ber88} are counted.
`Total' defines the 2016 database.}
\begin{tabular}{cccc}
\hline
\hline
\hspace{.4cm} Publication date \hspace{.6cm} & \hspace{.3cm}  No.\ of $pp$ data \hspace{.3cm} & \hspace{.3cm} No.\ of $np$ data \hspace{.3cm} & \hspace{.3cm} References \hspace{.2cm} \\
\hline
Jan.\ 1955 -- Dec.\ 1992 & 1787 & 2514 &   \cite{Ber90,Sto93} \\
Jan.\ 1993 -- Dec.\ 1999 & 1145 & 544 &   Tables XV and XVI of Ref.~\cite{Mac01} \\
Jan.\ 2000 -- Dec.\ 2016 & 140 & 511 &   Ref.~\cite{Al04} and Table~\ref{tab_npdata} of present paper \\
\hline
Total & 3072 & 3569 \\
\hline
\hline
\end{tabular}
\label{tab_NNhist}
\end{table}

\begin{table}
\caption{After-1999 $np$ data below 350 MeV included in the 2016 $np$ database.
``Error'' refers to the experimental over-all normalization errors of the individual datasets. `None' signifies that the respective experimental data set does not carry
a normalization error, i.e., the data are absolute. `Float' indicates that, in the analysis of the data set, the normalization was allowed to assume a value for which the $\chi^2$ is a minimum
disregarding a comparison with the experimental normalization error. This is done in case where there is doubt about the
alleged experimental normalization error. In the cases of `None' and `Float', the normalization is not counted as an observable.
This table contains 473 observables plus 38 normalizations resulting in a total of 511 data.
For the observables, we use in general the notation of Hoshizaki~\cite{Hos68}, except for
types which are undefined in the Hoshizaki formalism, where we use the Saclay notation~\cite{BLW78}.
}
\smallskip
\begin{tabular*}{\textwidth}{@{\extracolsep{\fill}}ccccc}
\hline 
\hline 
\noalign{\smallskip}
$T_{\rm lab}$ (MeV)    & \hspace{.3cm}  No. type  \hspace{.3cm} &  \hspace{.3cm} Error (\%)  \hspace{.3cm}  & \hspace{.3cm} Institution \hspace{.3cm} &  \hspace{.3cm}  Ref.   \hspace{.3cm} \\
\hline
\noalign{\smallskip}
9.2--349.0 & 92 $\sigma_{\rm tot}$ & None & Los Alamos & \cite{Ab01} \\
10.0 & 6 $\sigma$ & 0.8 & Ohio & \cite{Bo02} \\
95.0 & 10 $\sigma$ & 5.0 & Uppsala &\cite{Me04}\\
95.0 & 9 $\sigma$ & 4.0 & Uppsala & \cite{Me05}\\
96.0 & 11 $\sigma$ & 5.0 & Uppsala & \cite{Kl02} \\
96.0 & 9 $\sigma$ & 3.0 & Uppsala & \cite{Bl04} \\
96.0 & 12 $\sigma$ & None & Uppsala & \cite{Jo05} \\
260.0 & 8 $P$ & 1.8 & PSI & \cite{Ar00a}\\
260.0 & 16 $P$ & 1.8 & PSI & \cite{Ar00a}\\
260.0 & 8 $A_{yy}$ & 3.9 & PSI & \cite{Ar00a}\\
260.0 & 16 $A_{yy}$ & 3.9 & PSI & \cite{Ar00a}\\
260.0 & 9 $A_{zz}$ & 7.2 & PSI & \cite{Ar00a}\\
260.0 & 5 $D$ & 2.4 & PSI & \cite{Ar00b}\\
260.0 & 8 $D$ & Float & PSI & \cite{Ar00b}\\
260.0 & 8 $D_{0s''0k}$ & Float & PSI & \cite{Ar00b}\\
260.0 & 5 $D_t$ & 2.4 & PSI & \cite{Ar00b}\\
260.0 & 4 $A_t$ & 2.4 & PSI & \cite{Ar00b}\\
260.0 & 8 $A_t$ & 2.4 & PSI & \cite{Ar00b}\\
260.0 & 4 $R_t$ & 2.4 & PSI & \cite{Ar00b}\\
260.0 & 8 $R_t$ & 2.4 & PSI & \cite{Ar00b}\\
260.0 & 8 $N_{0nkk}$ & 2.4 & PSI & \cite{Ar00b}\\
260.0 & 4 $N_{0s''kn}$ & 2.4 & PSI & \cite{Ar00b}\\
260.0 & 8 $N_{0s''kn}$ & 2.4 & PSI & \cite{Ar00b}\\
260.0 & 4 $N_{0s''sn}$ & 2.4 & PSI & \cite{Ar00b}\\
260.0 & 8 $N_{0s''sn}$ & 2.4 & PSI & \cite{Ar00b}\\
284.0 & 14 $P$    &  3.0  & PSI &  \cite{Da02} \\
314.0 & 14 $P$    &  3.0  & PSI &  \cite{Da02} \\
315.0 & 16 $P$ & 1.2 & PSI & \cite{Ar00a}\\
315.0 & 11 $A_{yy}$ & 3.7 & PSI & \cite{Ar00a}\\
315.0 & 16 $A_{yy}$ & 3.7 & PSI & \cite{Ar00a}\\
315.0 & 11 $A_{zz}$ & 7.1 & PSI & \cite{Ar00a}\\
315.0 & 6 $D$ & Float & PSI & \cite{Ar00b}\\
315.0 & 6 $D_{0s''0k}$ & Float & PSI & \cite{Ar00b}\\
315.0 & 8 $D_{0s''0k}$ & Float & PSI & \cite{Ar00b}\\
315.0 & 6 $D_t$ & 1.9 & PSI & \cite{Ar00b}\\
315.0 & 6 $A_t$ & 1.9 & PSI & \cite{Ar00b}\\
315.0 & 8 $A_t$ & 1.9 & PSI & \cite{Ar00b}\\
315.0 & 6 $R_t$ & 1.9 & PSI & \cite{Ar00b}\\
315.0 & 8 $R_t$ & 1.9 & PSI & \cite{Ar00b}\\
315.0 & 5 $N_{0s''kn}$ & 1.9 & PSI & \cite{Ar00b}\\
315.0 & 8 $N_{0s''kn}$ & 1.9 & PSI & \cite{Ar00b}\\
315.0 & 6 $N_{0s''sn}$ & 1.9 & PSI & \cite{Ar00b}\\
315.0 & 8 $N_{0s''sn}$ & 1.9 & PSI & \cite{Ar00b}\\
315.0 & 8 $N_{0nkk}$ & 1.9 & PSI & \cite{Ar00b}\\
344.0 & 14 $P$    &  3.0  & PSI &  \cite{Da02} \\
\hline
\hline
\end{tabular*}
\label{tab_npdata}
\end{table}

\section{$NN$ scattering and the deuteron}
\label{sec_NN_scat}

Based upon the formalism presented in the previous section, we have constructed
$NN$ potentials through five orders of the chiral expansion, ranging from LO ($Q^0$) to N$^4$LO ($Q^5$).
In each order, we consider three cutoffs, namely, $\Lambda=$ 450, 500, and 550 MeV.
Since we take charge-dependence into account, each $NN$ potential comes in three versions:
$pp$, $np$, and $nn$.
The results from these potentials for $NN$ scattering and the deuteron will be presented in this
section.

\subsection{$NN$ database}
\label{sec_data}

Since an important part of $NN$ potential construction involves optimizing the reproduction
of the $NN$ data by the potential, we need to state, first, what $NN$ database we are using.

Our database consists of all $NN$ data below 350 MeV laboratory energy published in refereed physics journals between January 1955 and December 2016 that
are not discarded when applying the Nijmegen rejection criteria~\cite{Ber88}. We will refer to this as the
``2016 database''.
This  database was started by the Nijmegen group who critically checked and assembled the data published up to December 1992. This 1992 database consists of 1787 $pp$ data (listed in 
Ref.~\cite{Ber90}) and 2514 $np$ data (tabulated in Ref.~\cite{Sto93}), cf.\ Table~\ref{tab_NNhist}.
In Ref.~\cite{Mac01}, the database was then extended to include the data published up to December 1999 that survived the Nijmegen rejection criteria. This added 1145 $pp$ and 544 $np$ data (given in Tables XV and XVI of Ref.~\cite{Mac01}, respectively). Thus, the 1999 database includes
2932 $pp$ and 3058 $np$ data.

To get to the 2016 database, we have added to the 1999 database the data published
between January 2000 and December 2016 that are not rejected by the Nijmegen criteria.
We are aware of the fact that modified rejection criteria have been proposed~\cite{GS08} and applied
in recent $NN$ data analysis work~\cite{PAA13}.
But we continue to apply the classic Nijmegen criteria~\cite{Ber88} to be consistent with the pre-2000 part of the database.

Concerning after-1999 $pp$ data, there exists only one set of 139 differential cross sections between 239.9 and 336.2 MeV measured by the EDDA group at COSY (J\H{u}lich, Germany) with an over-all uncertainty of 2.5\%~\cite{Al04}. Thus, the total number of $pp$ data contained in the 2016 database is 3072 (Table~\ref{tab_NNhist}).

In contrast to $pp$, there have been many new $np$ measurements after 1999. We list the datasets
that survived the Nijmegen rejection criteria in Table~\ref{tab_npdata}. 
According to that list, the number of valid after-1999 $np$
data is 511, bringing the total number of $np$ data contained in the 2016 database to 
3569  (Table~\ref{tab_NNhist}).

For comparison, we mention that the 2013 Granada $NN$ database~\cite{PAA13} consists of 2996 $pp$
and 3717 $np$ data. The larger number of $pp$ data in our base is mainly due to
the inclusion of 140 $pp$ data from Ref.~\cite{Al04} which are left out in the Granada base.
On the other hand, the Granada base contains 148 more $np$ data which is
a consequence of the modified rejection criteria applied by the Granada group which allows for the survival of more $np$ data. 

Finally, we note that in the potential construction reported in this paper, we make use of the 2016 database only up to 290 MeV laboratory energy (pion-production threshold).
Between 0 and 290 MeV, the 2016 database contains 2132 $pp$ data and 2721 $np$ data
(cf.\ Table~\ref{tab_chi}).

\begin{figure}[t]
\vspace*{-0.5cm}
\hspace*{-0.7cm}
\scalebox{0.45}{\includegraphics{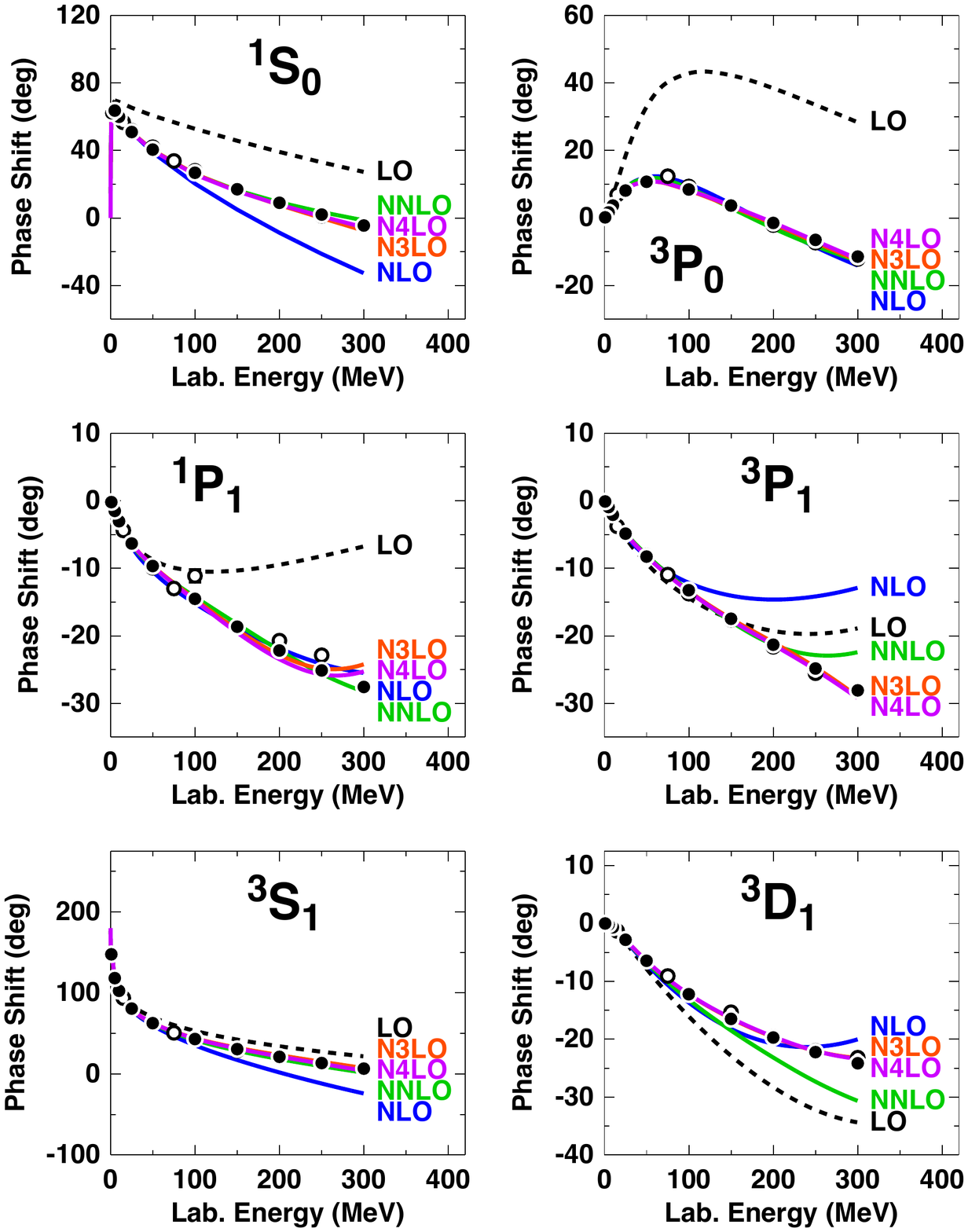}}
\hspace*{-1.2cm}
\scalebox{0.45}{\includegraphics{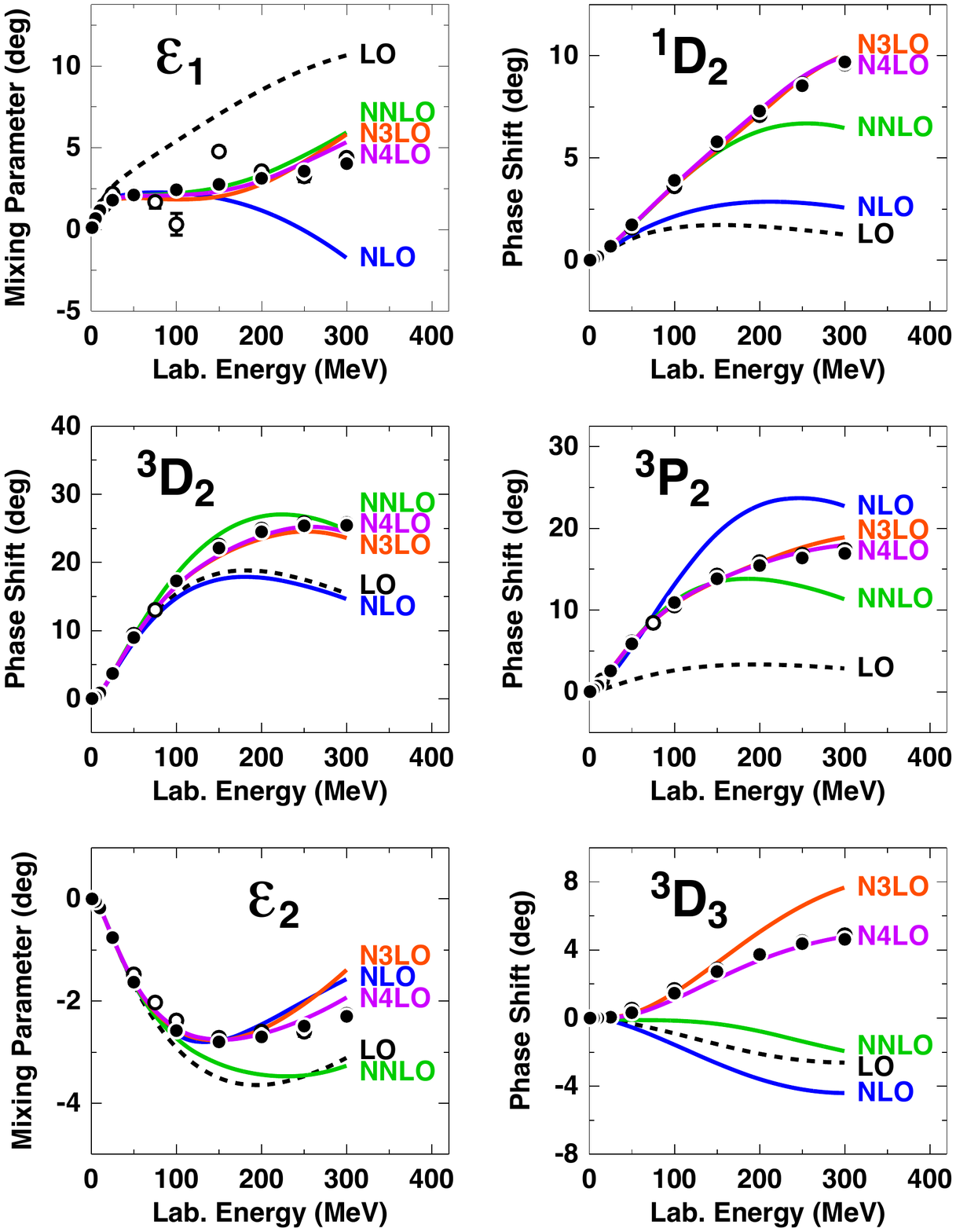}}
\vspace*{-0.5cm}
\caption{(Color online).
Chiral expansion of neutron-proton scattering as represented by the phase shifts in 
$S$, $P$, and $D$ waves and mixing parameters 
$\epsilon_1$ and $\epsilon_2$. Five orders ranging from LO to N$^4$LO are shown as denoted.
A cutoff $\Lambda=500$ MeV is applied in all cases.
The filled and open circles represent the results from the Nijmegen multi-energy $np$ phase-shift analysis~\cite{Sto93} and the GWU single-energy $np$ analysis SP07~\cite{SP07}, respectively.
\label{fig_figph1}}
\end{figure}

\begin{table}
\caption{$\chi^2/$datum for the fit of the 2016 $NN$ data base by $NN$ potentials at various orders of chiral EFT ($\Lambda = 500$ MeV in all cases).
\label{tab_chi}}
\smallskip
\begin{tabular*}{\textwidth}{@{\extracolsep{\fill}}ccccccc}
\hline 
\hline 
\noalign{\smallskip}
 $T_{\rm lab}$ bin (MeV) & No.\ of data & LO & NLO & NNLO & N$^3$LO & N$^4$LO \\
\hline
\noalign{\smallskip}
\multicolumn{7}{c}{\bf proton-proton} \\
0--100 & 795 & 520 & 18.9  & 2.28   &  1.18 & 1.09 \\
0--190 & 1206  & 430 & 43.6  &  4.64 & 1.69 & 1.12 \\
0--290 & 2132 & 360 & 70.8  & 7.60  &  2.09  & 1.21 \\
\hline
\noalign{\smallskip}
\multicolumn{7}{c}{\bf neutron-proton} \\
0--100 & 1180 & 114 & 7.2  &  1.38  & 0.93  & 0.94 \\
0--190 & 1697 &  96 & 23.1  &  2.29 &  1.10  & 1.06 \\
0--290 & 2721 &  94 &  36.7 & 5.28  &  1.27 & 1.10 \\
\hline
\noalign{\smallskip}
\multicolumn{7}{c}{\boldmath $pp$ plus $np$} \\
0--100 & 1975 & 283 &  11.9 &  1.74  & 1.03   & 1.00  \\
0--190 & 2903 & 235 &  31.6 &  3.27 & 1.35   &  1.08 \\
0--290 & 4853 & 206 &  51.5 & 6.30  & 1.63   &  1.15 \\
\hline
\hline
\noalign{\smallskip}
\end{tabular*}
\end{table}

\begin{table}
\caption{Scattering lengths ($a$) and effective ranges ($r$) in units of fm as predicted by $NN$ potentials at various orders of chiral EFT ($\Lambda = 500$ MeV in all cases).
($a_{pp}^C$ and $r_{pp}^C$ refer to the $pp$ parameters in the presence of
the Coulomb force. $a^N$ and $r^N$ denote parameters determined from the
nuclear force only and with all electromagnetic effects omitted.)
$a_{nn}^N$, and $a_{np}$ are fitted, all other quantities are predictions.
\label{tab_lep}}
\smallskip
\begin{tabular*}{\textwidth}{@{\extracolsep{\fill}}ccccccc}
\hline 
\hline 
\noalign{\smallskip}
 & LO & NLO & NNLO & N$^3$LO & N$^4$LO & Empirical \\
\hline
\noalign{\smallskip}
\multicolumn{7}{c}{\boldmath $^1S_0$} \\
$a_{pp}^C$  &--7.8153 & --7.8128& --7.8140 & --7.8155  & --7.8160
  &--7.8196(26)~\cite{Ber88} \\
  &&&&&& --7.8149(29)~\cite{SES83} \\
$r_{pp}^C$  & 1.886   & 2.678 & 2.758 & 2.772 & 2.774
  & 2.790(14)~\cite{Ber88}  \\
  &&&&&& 2.769(14)~\cite{SES83} \\
$a_{pp}^N$  &--- & --17.476 & --17.762 & --17.052 & --17.123 & --- \\
$r_{pp}^N$  &  ---  & 2.752& 2.821 & 2.851 & 2.853 & --- \\
$a_{nn}^N$  &--18.950 & --18.950& --18.950 & --18.950 & --18.950
&--18.95(40)~\cite{Gon06,Che08} \\
$r_{nn}^N$  &  1.857  & 2.726 & 2.800 & 2.812 &  2.816 &  2.75(11)~\cite{MNS90} \\
$a_{np}  $  &--23.738 & --23.738 & --23.738 & --23.738 & --23.738
 &--23.740(20)~\cite{Mac01} \\
$r_{np}  $  &  1.764  & 2.620 & 2.687 & 2.700 & 2.704 & [2.77(5)]~\cite{Mac01}   \\
\hline
\noalign{\smallskip}
\multicolumn{7}{c}{\boldmath $^3S_1$} \\
$a_t$     &  5.255   & 5.415 & 5.418 & 5.420 & 5.420 & 5.419(7)~\cite{Mac01}  \\
$r_t$     &  1.521   & 1.755 & 1.752 & 1.754 & 1.753 & 1.753(8)~\cite{Mac01}  \\
\hline
\hline
\noalign{\smallskip}
\end{tabular*}
\end{table}

\begin{table}
\small
\caption{Two- and three-nucleon bound-state properties as predicted by
  $NN$ potentials at various orders of chiral EFT ($\Lambda = 500$ MeV in all cases).
(Deuteron: Binding energy $B_d$, asymptotic $S$ state $A_S$,
asymptotic $D/S$ state $\eta$, structure radius $r_{\rm str}$,
quadrupole moment $Q$, $D$-state probability $P_D$; the predicted
$r_{\rm str}$ and $Q$ are without meson-exchange current contributions
and relativistic corrections. Triton: Binding energy $B_t$.)
$B_d$ is fitted, all other quantities are predictions.
\label{tab_deu}}
\smallskip
\begin{tabular*}{\textwidth}{@{\extracolsep{\fill}}lllllll}
\hline 
\hline 
\noalign{\smallskip}
 & LO & NLO & NNLO & N$^3$LO & N$^4$LO & Empirical$^a$ \\
\hline
\noalign{\smallskip}
{\bf Deuteron} \\
$B_d$ (MeV) &
 2.224575& 2.224575 &
 2.224575 & 2.224575 & 2.224575 & 2.224575(9) \\
$A_S$ (fm$^{-1/2}$) &
 0.8526& 0.8828 &
0.8844 & 0.8853 & 0.8852 & 0.8846(9)  \\
$\eta$         & 
 0.0302& 0.0262 &
0.0257& 0.0257 & 0.0258 & 0.0256(4) \\
$r_{\rm str}$ (fm)   & 1.911
      & 1.971 & 1.968
       & 1.970
       & 1.973 &
 1.97507(78) \\
$Q$ (fm$^2$) &
 0.310& 0.273&
 0.273 & 
 0.271 & 0.273 &
 0.2859(3)  \\
$P_D$ (\%)    & 
 7.29& 3.40&
4.49 & 4.15 & 4.10 & --- \\
\hline
\noalign{\smallskip}
{\bf Triton} \\
$B_t$ (MeV) & 11.09  & 8.31  & 8.21 & 8.09  & 8.08 & 8.48 \\
\hline
\hline
\noalign{\smallskip}
\end{tabular*}
\footnotesize
$^a$See Table XVIII of Ref.~\cite{Mac01} for references;
the empirical value for $r_{\rm str}$ is from Ref.~\cite{Jen11}.\\
\end{table}

\subsection{Data fitting procedure}
\label{sec_fitting}

When we are talking about data fitting, we are referring to the adjustment of the $NN$ contact parameters available at the respective order. Note that in our $NN$ potential construction, the $\pi N$ LECs are not fit-parameters. The $\pi N$ LECs are held fixed at their values determined in the $\pi N$ analysis of Ref.~\cite{Hof15} displayed in Table~\ref{tab_lecs} (we use the central values shown in that Table). Thus, the $NN$ contacts (Sec.~\ref{sec_short}) are the only fit parameters used to optimize the reproduction of the $NN$ data below 290 MeV laboratory energy.
As discussed, those contact terms describe the short-range part of the $NN$ potentials and adjust the lower partial waves.

In the construction of any $NN$ potential, we always start with the $pp$ version since the $pp$ data are the most accurate ones. The fitting is done in three steps. In the first step, the $pp$ potential
is adjusted to reproduce as closely as possible the $pp$ phase shifts of the Nijmegen multienergy $pp$ phase shift analysis~\cite{Sto93} up to 300 MeV laboratory energy. This is to ensure that phase shifts are in the right ballpark.
In the second step, we make use of the Nijmegen $pp$ error matrix~\cite{SS93} to minimize the $\chi^2$ that results from it. The advantage of this step is that it is computationally very fast and easy.
Finally, in the third and final step, the $pp$ potential contact parameters are fine-tuned by minimizing the $\chi^2$ that results from a direct comparison with the experimental $pp$ data
contained in the 2016 database below 290 MeV. For this we use a copy of the SAID software
package which includes  all electromagnetic contributions necessary for the calculation of $NN$ observables at low energy. Since it turned out that the Nijmegen error matrix produces very accurate $\chi^2$ for $pp$ energies below 75 MeV, we use the values from this error matrix for the energies up to 75 MeV and the values from a direct confrontation with the data above that energy.

The $I=1$ $np$ potential is constructed by starting from the $pp$ version, applying the charge-dependence discussed in Sec.~\ref{sec_CD}, and adjusting the non-derivative $^1S_0$ contact such as to reproduce the $^1S_0$ $np$ scattering length. This then yields the preliminary fit of
the $I=1$ $np$ potential. 
The preliminary fit of the $I=0$ $np$ potential is obtained by a fit to 
the $I=0$ $np$ phase shifts of the Nijmegen multienergy $np$ phase shift analysis~\cite{Sto93} below 300 MeV.
Starting from this preliminary $np$ fit, the contact parameters are fine-tuned in a confrontation with the $np$ data below 290 MeV, for which the $\chi^2$ is minimized. We note that during this last step we have also allowed for minor changes of the $I=1$ parameters (which also modifies the $pp$ potential) to obtain an even lower $\chi^2$ over-all. 

Finally the $nn$ potential is obtained by starting from the $pp$ version, replacing the proton masses by neutron masses, leaving out Coulomb, and adjusting the 
non-derivative $^1S_0$ contact such as to reproduce the $^1S_0$ $nn$ scattering length
for which we assume the empirical value of $-18.95$ MeV\cite{Gon06,Che08}.

We note that our procedure for fitting $NN$ potentials to data is essentially the same  that was used to fit the 
high-precision $NN$ potentials of the 1990's~\cite{Sto94,WSS95,Mac01} (fitted up to 350 MeV),
and the first precision chiral $NN$ potentials~\cite{EM03,ME11} (fitted up to 290 MeV).
This is quite in contrast to the procedure applied in the recent construction of the  NNLO$_{\rm sat}$ potential~\cite{Eks15}, where the $NN$ data up to 35 MeV and the groundstate energies and  radii of nuclei up to
$^{40}$Ca are taken into acount to fix simulteneously the 2NF and 3NF. In Ref.~\cite{Eks15}, the $NN$ data up to 35 MeV are reproduced with a $\chi^2$/datum of 4.3. Similar procedures are applied in Ref.~\cite{Car16}.

Our fit procedures differ also substatially from the ones used in the recent chiral $NN$ potential constructions of Refs~\cite{EKM15,EKM15a}, where the potentials are fitted to phase shifts. Already in the early 1990's, the Nijmegen group has pointed out repeatedly and demonstrated clearly~\cite{SS93}, that fitting to experimental data should be preferred over fitting to phase shifts, 
because a seemingly ``good'' fit to phase shifts
can result in a bad reproduction of the data. Note that phase shifts are not experimental data.

\subsection{Results for $NN$ scattering}
\label{sec_results}

The $\chi^2$/datum for the reproduction of the $NN$ data at various orders of chiral EFT 
are shown in Table~\ref{tab_chi} for different energy intervals below 290 MeV laboratory energy ($T_{\rm lab}$). 
The bottom line of Table~\ref{tab_chi} summarizes the essential results.
For the close to 5000 $pp$ plus $np$ data below 290 MeV (pion-production threshold),
the $\chi^2$/datum 
is 51.4 at NLO and 6.3 at NNLO. Note that the number of $NN$ contact terms is the same for both orders. The improvement is entirely due to an improved description of the 2PE contribution, which is responsible for the crucial intermediate-range attraction of the nuclear force.
At NLO, only the uncorrelated 2PE is taken into account which is insufficient. From the classic meson-theory of nuclear forces~\cite{MHE87}, it is wellknown that $\pi$-$\pi$ correlations and nucleon resonances need to be taken into account for a realistic model of 2PE that provides a sufficient
amount of intermediate attraction to properly bind nucleons in nuclei.  In the chiral theory, these contributions are encoded in the subleading $\pi N$ vertexes with LECs denoted by $c_i$.
These enter at NNLO and are the reason for the substantial improvements we encounter at that order.
This is the best proof that, starting at NNLO, the chiral approach to nuclear forces is getting the physics right. 

To continue on the bottom line of Table~\ref{tab_chi}, after NNLO,
the $\chi^2$/datum then further improves to 1.63 at N$^3$LO and, finally, reaches the almost perfect value of 1.15 at N$^4$LO---a fantastic convergence.

Corresponding $np$ phase shifts are displayed in Fig.~\ref{fig_figph1}, which reflect what
the $\chi^2$ have already proven, namely, an excellent convergence when going from NNLO to N$^3$LO and, finally, to N$^4$LO. 
However, at LO and NLO there are large discrepancies between the predictions and
the empirical phase shifts as to be expected from the corresponding $\chi^2$ values.
 This fact renders applications of the LO and NLO nuclear force useless for any realistic calculation (but they could be used to demonstrate truncation errors).

For order N$^4$LO (with $\Lambda=500$ MeV), we also provide the numerical values
for the phase shifts in Appendix~\ref{app_ph}. Our $pp$ phase shifts 
are the phase shifts of the nuclear plus relativistic Coulomb interaction with respect
to Coulomb wave functions. Note, however, that for the calculation of observables
(e.g., to obtain the $\chi^2$ in regard to experimental data),
we use electromagnetic phase shifts, as {\it necessary}, which we obtain by adding to the Coulomb phase shifts the effects from two-photon exchange, vacuum polarization, and magnetic moment interactions as calculated by the Nijmegen group~\cite{Ber88,Sto95}.
This is important for $^1S_0$ below 30 MeV and negligible otherwise. 
For $nn$ and $np$ scattering, our phase shifts are the ones from the nuclear interaction
with respect to Riccati-Bessel functions. The technical details of our phase shift calculations
can be found in appendix A3 of Ref.~\cite{Mac01}.

The low-energy scattering parameters, order by order,  are shown in Table~\ref{tab_lep}.
For $nn$ and $np$, the effective range expansion without any electromagnetic interaction is used.
In the case of $pp$ scattering, the quantities $a_{pp}^C$ and $r_{pp}^C$
are obtained by using the effective range expansion appropriate in the presence of the
Coulomb force (cf.\ appendix A4 of Ref.~\cite{Mac01}). Note that the empirical values for
$a_{pp}^C$ and $r_{pp}^C$
in Table~\ref{tab_lep} were obtained by subtracting from the corresponding electromagnetic values the effects due to two-photon exchange and vacuum polarization. Thus, the comparison between theory and experiment for these two quantities is conducted correctly.
$a_{nn}^N$, and $a_{np}$ are fitted, all other quantities are predictions.
Note that the $^3S_1$ effective range parameters $a_t$ and $r_t$ are not fitted.
But the deuteron binding energy is fitted (cf.\ next subsection) and that essentially fixes
$a_t$ and $r_t$.

\subsection{Deuteron and triton}
\label{sec_deu}

The evolution of the deuteron properties from LO to N$^4$LO  of chiral EFT are shown in Table~\ref{tab_deu}.
In all cases, we fit the deuteron binding energy to its empirical value of 2.224575 MeV
using the non-derivative $^3S_1$ contact. All other deuteron properties are predictions.
Already at NNLO, the deuteron has converged to its empirical properties and stays there
through the higher orders.

At the bottom of Table~\ref{tab_deu}, we also show the predictions for the triton binding
as obtained in 34-channel charge-dependent Faddeev calculations using only 2NFs. The results show smooth and steady convergence, order by order, towards a value around 8.1 MeV that is reached at the highest orders shown. This contribution from the 2NF will require only a moderate 3NF. 
The relatively low deuteron $D$-state probabilities ($\approx 4.1$\% at N$^3$LO and N$^4$LO) and the concomitant generous triton binding energy predictions are
a reflection of the fact that our $NN$ potentials are soft (which is, at least in part, due to their non-local character).

\subsection{Cutoff variations}

\begin{figure}[th]
\vspace*{-0.5cm}
\hspace*{-0.7cm}
\scalebox{0.45}{\includegraphics{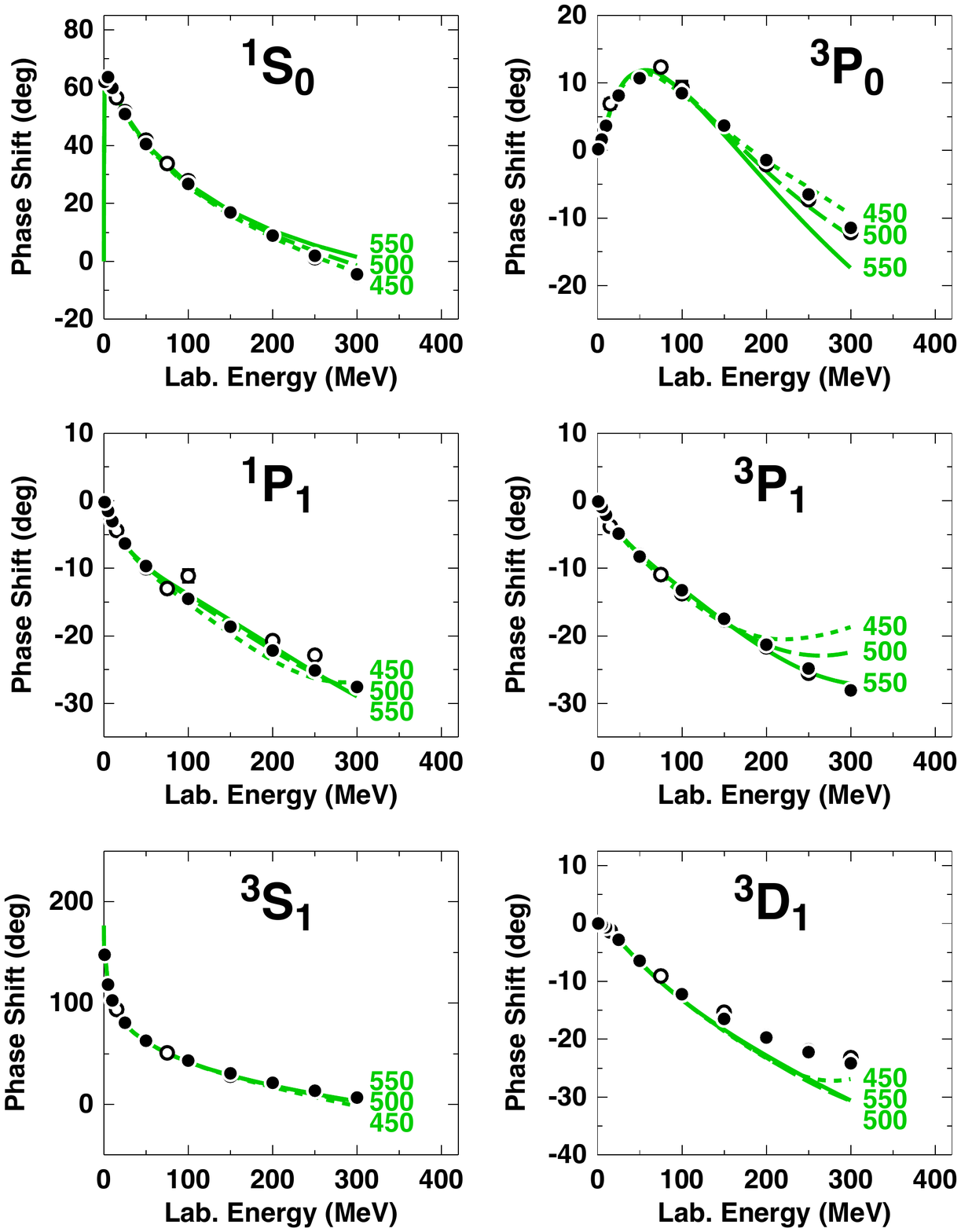}}
\hspace*{-1.2cm}
\scalebox{0.45}{\includegraphics{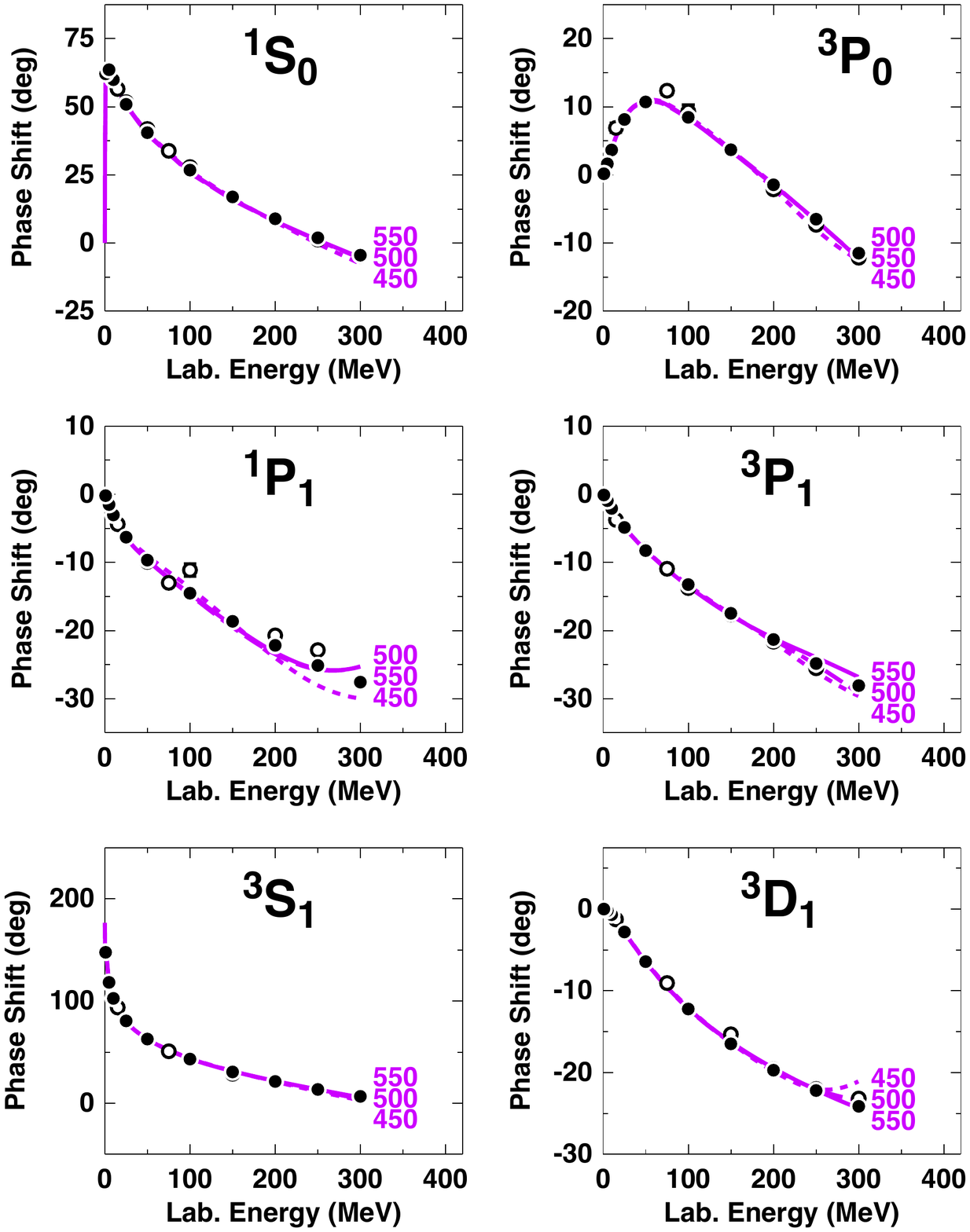}}

\vspace*{-1.5cm}
\hspace*{-0.7cm}
\scalebox{0.45}{\includegraphics{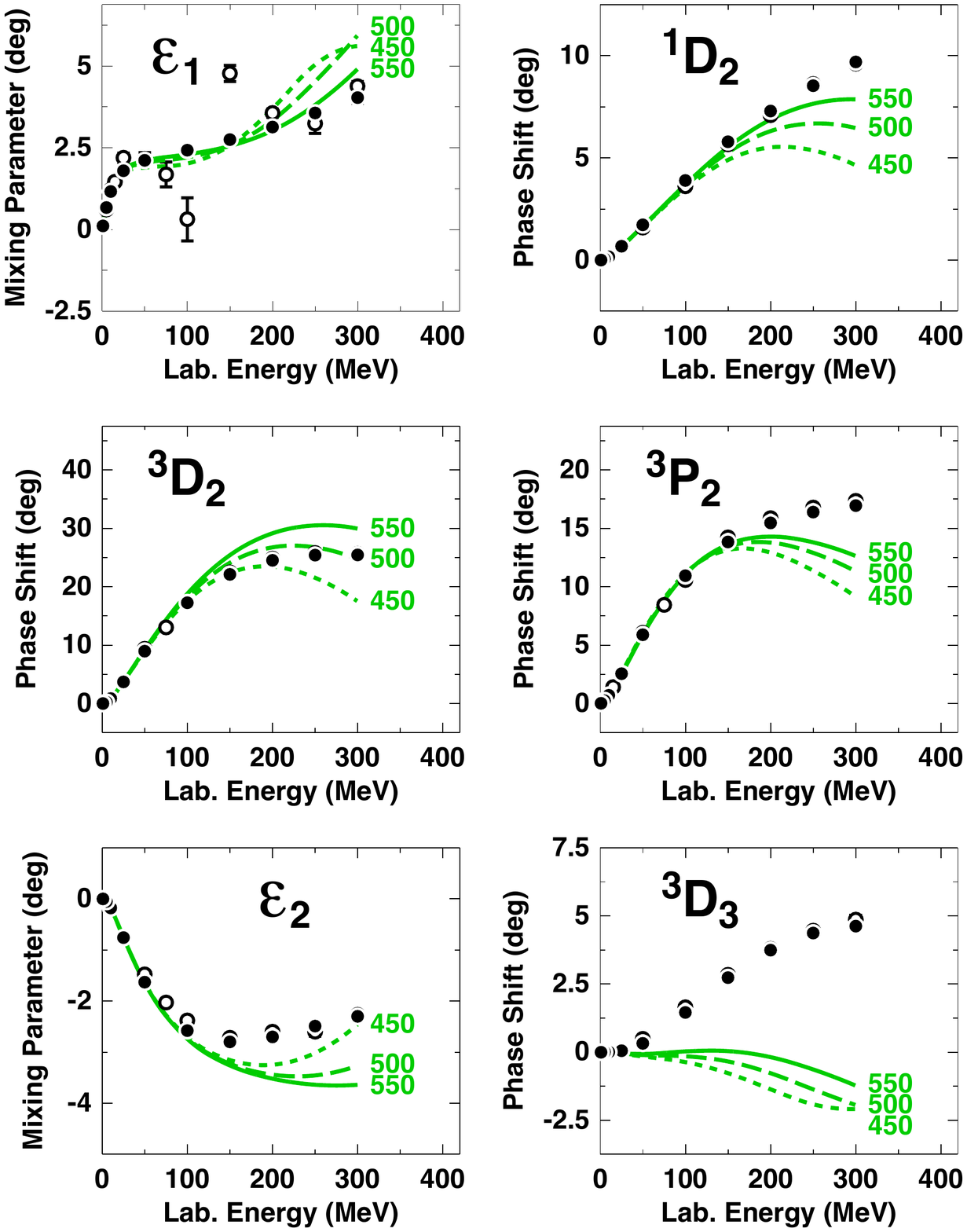}}
\hspace*{-1.2cm}
\scalebox{0.45}{\includegraphics{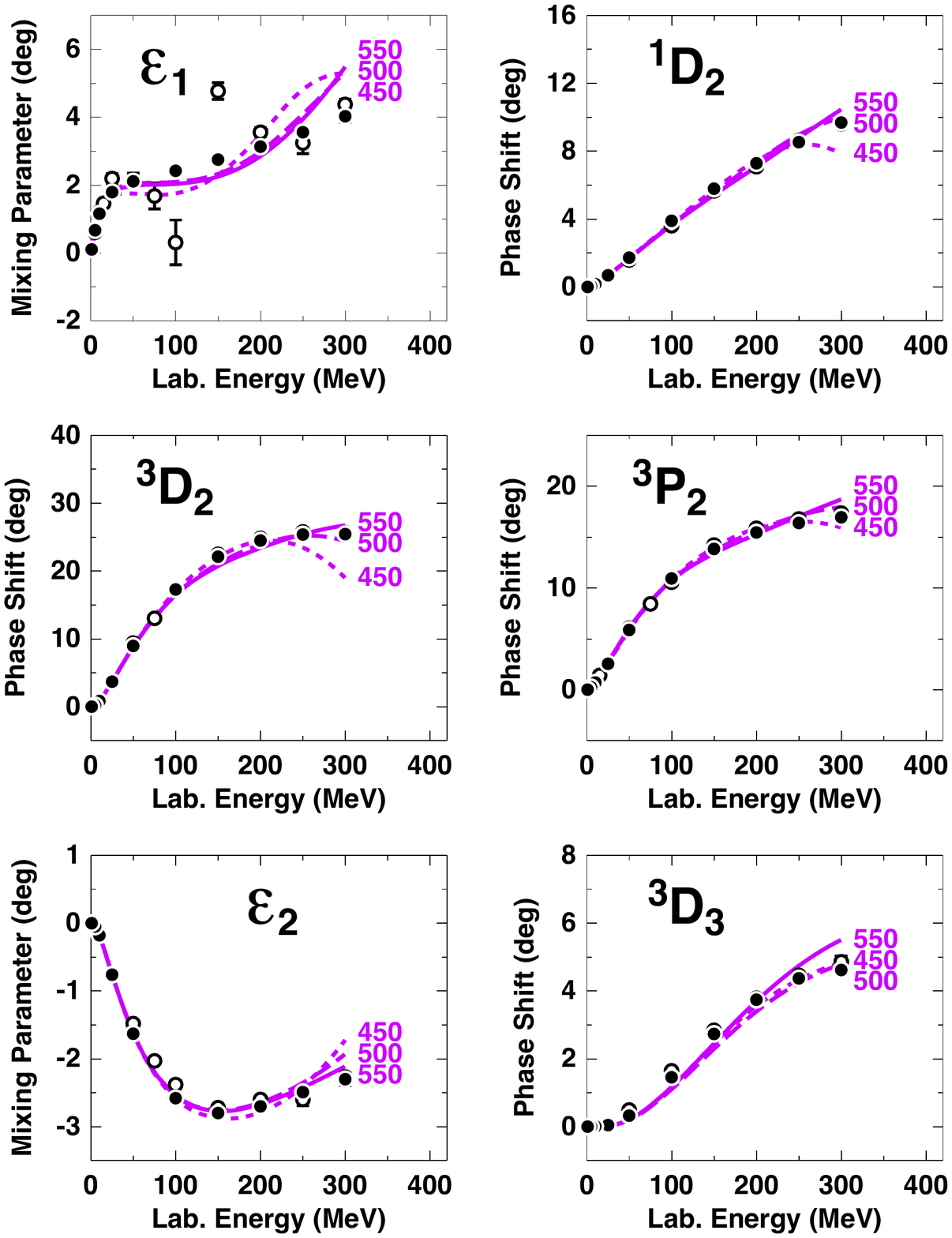}}
\vspace*{-1.0cm}
\caption{(Color online).
Cutoff variations of the $np$ phase shifts at NNLO (left side, green lines) and N$^4$LO (right side, purple lines). Dotted, dashed, and solid lines represent the results obtained with cutoff parameter $\Lambda =$ 450, 500, and 550 MeV, respectively, as also indicated by the curve labels.
Note that, at N$^4$LO, the cases 500 and 550 MeV cannot be distinguished on the scale of the figures for most partial waves.
Filled and open circles as in Fig.~\ref{fig_figph1}.
\label{fig_figph2}}
\end{figure}

\begin{table}
\small
\caption{$\chi^2$/datum for for the fit of the $pp$ plus $np$ data up to 190 MeV and two- and three-nucleon bound-state properties as produced by
  $NN$ potentials at NNLO and N$^4$LO applying different values for the cutoff
  parameter $\Lambda$ of the regulator function Eq.~(\ref{eq_f}).
  For some of the notation, see Table~\ref{tab_deu}, where also empirical information on the deuteron and triton can be found.
\label{tab_deu2}}
\smallskip
\begin{tabular*}{\textwidth}{@{\extracolsep{\fill}}lllllll}
\hline 
\hline 
\noalign{\smallskip}
  & \multicolumn{3}{c}{NNLO} & \multicolumn{3}{c}{N$^4$LO} \\
               \cline{2-4}                            \cline{5-7}
  $\Lambda$(MeV) & 450 & 500 & 550 & 450 & 500 & 550 \\ 
  \hline
\noalign{\smallskip}
{\bf\boldmath $\chi^2$/datum $pp$ \& $np$} \\
0--190 MeV (2903 data) &
 4.12 & 3.27 & 3.32 &
 1.17 & 1.08 & 1.25 \\
\hline
\noalign{\smallskip}
{\bf Deuteron} \\
$B_d$ (MeV) &
 2.224575& 2.224575&
 2.224575 & 2.224575 & 2.224575 & 2.224575 \\
$A_S$ (fm$^{-1/2}$) &
0.8847 & 0.8844 & 0.8843 & 
 0.8852 & 0.8852 & 0.8851  \\
$\eta$         & 
0.0255 & 0.0257 & 0.0258 &
 0.0254 & 0.0258 & 0.0257 \\
$r_{\rm str}$ (fm)   & 
    1.967 & 1.968 & 1.968 &
   1.966 & 1.973 & 1.971 \\
$Q$ (fm$^2$) &
0.269  & 0.273 & 0.275 & 
 0.269 & 0.273 & 0.271  \\
$P_D$ (\%)    & 
 3.95 & 4.49 & 4.87 &
4.38 & 4.10 & 4.13 \\
\hline
\noalign{\smallskip}
{\bf Triton} \\
$B_t$ (MeV) & 
 8.35 & 8.21 & 8.10 & 
8.04  & 8.08 & 8.12 \\
\hline
\hline
\noalign{\smallskip}
\end{tabular*}
\end{table}

As noted before, besides the case $\Lambda=500$ MeV, we have also constructed potentials with 
$\Lambda=450$ and 550 MeV at each order, to allow for systematic
studies of the cutoff dependence.
In Fig.~\ref{fig_figph2}, we display the variations of the $np$ phase shifts for different cutoffs at NNLO (left half of figure, green curves) and at N$^4$LO (right half of figure, purple curves).
We do not show the cutoff variations of phase shifts at N$^3$LO, because they are about the same as at N$^4$LO. Similarly, the variations at NLO are of about the same size as at NNLO.
Fig.~\ref{fig_figph2} demonstrates nicely how cutoff dependence
diminishes with increasing order---a reasonable trend.
Another point that is evident from this figure is that $\Lambda=450$ MeV should be considered
as a lower limit for cutoffs, because obviously cutoff artifacts start showing up---above 200 MeV, particularly,
in $^1D_2$ and $^3D_2$.
Concerning the upper limit for the cutoff: It has been discussed and demonstrated in length in the literature 
(see, e.g., Ref.~\cite{EKM15}) that for the $NN$ interaction the breakdown scale occurs around 
$\Lambda_b \approx 600$ MeV.
The motivation for our upper value of 550 MeV is to stay below  $\Lambda_b$.

In Table~\ref{tab_deu2}, we show the cutoff dependence for three selected aspects that are of great interest: the $\chi^2$ for the fit of the $NN$ data below 190 MeV, the deuteron properties, and the triton binding energy. The $\chi^2$ does not change substantially as a function of cutoff, and
crucial deuteron properties, like $A_S$ and $\eta$, stay within the empirical range, for both NNLO and
N$^4$LO. Thus, we can make the interesting observation that the reproduction of $NN$ observables
is not much affected by the cutoff variations.
However, the $D$-state probability of the deuteron, $P_D$, which is not an observable, changes
substantially as a function of cutoff at NNLO (namely, by $\approx 1$\%) while it changes only
by 0.25\% at N$^4$LO. Note that $P_D$ is intimately related to the off-shell behavior of a potential
and so are the binding energies of few-body systems. Therefore, in tune with the $P_D$ variations,
the binding energy of the triton varies by 0.25 MeV at NNLO, while it 
changes only by 0.08 MeV at N$^4$LO.
In this context, it is of interest to note that changes in the off-shell behavior of the 2NF can be compensated by corresponding changes in the 3NF, as demonstrated by 
Polyzou and Gl\H{o}cke~\cite{PG90}.

Even though cutoff variations are, in general, not the most reliable way to estimate truncation errors, in the above case they seem to reflect closely what we expect to be the truncation error.

\section{Chiral three-body forces}
\label{sec_3nf}

\begin{table}
\caption{Effective $\pi N$ LECs (in units of GeV$^{-1}$) recommended for the 2PE 3NF at the given orders. See text for explanations.}
\label{tab_lecs_3nf}
\smallskip
\begin{tabular*}{\textwidth}{@{\extracolsep{\fill}}crrr}
\hline 
\hline 
\noalign{\smallskip}
              & NNLO & N$^3$LO & N$^4$LO \\
\hline
\noalign{\smallskip}
$\bar{c}_1$ & --0.74 & --1.20  & --0.73 \\
$\bar{c}_3$ & --3.61 & --4.43 & --3.38 \\
$\bar{c}_4$ & 2.44 & 2.67 & 1.69 \\
\hline
\hline
\noalign{\smallskip}
\end{tabular*}
\end{table}

As is well established, realistic {\it ab initio} nuclear structure calculations
require the inclusion of 3NFs (and potentially also four-nucleon forces).
The first 3NFs occur at NNLO (cf.\ Fig.~\ref{fig_hi}) and were derived in
Refs.~\cite{Kol94,Epe02}.
The 3NFs at N$^3$LO can be found in Refs.~\cite{Ber08,Ber11}.
Finally, at N$^4$LO, the longest-range and intermediate-range 3NFs are
given in Refs.~\cite{KGE12,KGE13}. 
Moreover, a new set of ten 3NF contact terms occurs at N$^4$LO, which has 
been derived by the Pisa group~\cite{GKV11}.
An efficient approach for calculating the matrix elements of chiral 3NF
contributions up to N$^3$LO has been published in
Ref.~\cite{Heb15}. This approach may eventually be extended to N$^4$LO.

In the derivation of all of the above-cited chiral 3NFs, the same power-counting scheme
is applied as in the derivation of the 2NFs of this paper, namely, 
Weinberg counting and considering $Q/M_N \sim (Q/\Lambda_\chi)^2$ (Sec.~\ref{sec_pow}).
Thus, those 3NF expressions are consistent with the present 2NFs, and they can be used 
together in {\it ab initio} calculations of nuclear structure and reactions.

In this context it is worth noting that, for convenience, the 3NFs are derived 
using dimensional regularization (DR), while we use SFR in the construction of the 2NFs [cf.~Eq.~(\ref{eq_disp})].
This is, however, not a problem because, as shown in Ref.~\cite{EGM04},
DR and SFR expressions differ only by higher order terms that are beyond
the given order. Thus, the accuracy of the calculation conducted at a given order is
not affected. An equivalent argument applies to the use of nonlocal regulators 
[Eq.~(\ref{eq_f})] {\it versus} local ones (e.g., Eq.~(11) of Ref.~\cite{Nav07}), since
also these two types of regulators differ only by higher order terms beyond
the given order.

Because of the complexity of the N$^4$LO 3NF, it may still take a few years until this force
is available in a manageable form.
Thus, for a while, we will have to live with incomplete calculations.
However, there is one important component of the 3NF where, indeed, complete
calculations up to N$^4$LO are possible: it is the 2PE 3NF. In Ref.~\cite{KGE12} it has been
shown that the 2PE 3NF has essentially the same mathematical structure at NNLO, N$^3$LO, and N$^4$LO.
Thus, one can add up the three orders of 3NF contributions and parametrize the result in terms of effective LECs. This was done in Ref.~\cite{KGE12} and we show the effectice LECs they come up with in Table~\ref{tab_lecs_3nf}, column N$^4$LO, where we quote the numbers given in Eq.~(5.2) of Ref.~\cite{KGE12}, which are 
based upon the GW $\pi N$ phase shifts~\cite{Arn06}. Note that the LECs of Ref.~\cite{Hof15}, which we are using for the 2NF, are
also based upon GW input. Thus, there is consistency between the effective $\bar{c}_i$ for the 3NF
(column N$^4$LO of  Table~\ref{tab_lecs_3nf}) and our $c_i$ for the 2NF
(column N$^4$LO of  Table~\ref{tab_lecs}).

Concerning, the 2PE 3NF at N$^3$LO, Eq.~(2.8) of Ref.~\cite{Ber08} provides the corrections to the $c_i$
when the 2PE 3NF at N$^3$LO is added in. 
Note, however, that there is a error in the numerical values given below Eq.~(2.8) of Ref.~\cite{Ber08}.
While $\delta c_1 = -0.13$ GeV$^{-1}$ is correct, the correct values for $\delta c_3$ and
 $\delta c_4$ are
$\delta c_3 = -\delta c_4 = 0.89$ GeV$^{-1}$.
 When these corrections are applied to the N$^3$LO $c_i$ of our Table~\ref{tab_lecs}, then the values given in the N$^3$LO column  of Table~\ref{tab_lecs_3nf} emerge.
By using the $\bar{c}_i$ of Table~\ref{tab_lecs_3nf} in the mathematical expression of the NNLO 3NF, one can include at least the 2PE parts of the 3NF up to N$^3$LO and
even up to N$^4$LO in a straightforward way.

The 2PE 3NF is the most obvious among all possible 3NF contributions. Historically, it is
the first 3NF ever calculated~\cite{FM57}. The above-given prescriptions allow to take
care of this very basic 3NF up to the highest orders considered in this paper.

\section{Uncertainty quantifications}
\label{sec_uncert}

When applying chiral two- and many-body forces in {\it ab initio} calculations producing predictions for observables of nuclear structure and reactions, 
major sources of uncertainties are~\cite{FPW15}:
\begin{enumerate}
\item
Experimental errors of the input $NN$ data that the 2NFs are based upon and the input
few-nucleon data to which the 3NFs are adjusted.
\item
Uncertainties in the Hamiltonian due to 
       \begin{enumerate}
       \item
       uncertainties in the determination of the $NN$ and $3N$ contact LECs,
       \item
       uncertainties in the $\pi N$ LECs, 
       \item
       regulator dependence, 
       \item
       EFT truncation error.
       \end{enumerate}
\item
Uncertainties associated with the few- and many-body methods applied.
\end{enumerate}

The experimental errors in the $NN$ scattering and deuteron data propagate into the                     
$NN$ potentials that are adjusted to reproduce those data.
 To systematically investigate this error propagation,
the Granada group has constructed smooth local
potentials~\cite{PAA14}, the parameters of which carry the uncertainties implied by the errors in the $NN$ data. 
Applying 205 Monte Carlo samples of these potentials, they find an uncertainty of 15 keV 
for the triton binding energy~\cite{Per14}.
In a more recent study~\cite{Per15}, in which only 33 Monte Carlo samples were used, 
the Granada group reproduced the uncertainty of 15 keV 
for the triton binding energy and, in addition, determined the uncertainty
for the $^4$He binding energy to be 55 keV.
The conclusion is that the statistical error propagation from the $NN$ input data to
the binding energies of light nuclei is negligible as compared to uncertainties from other sources (discussed below). Thus, this source of error can be safely neglected at this time.
Furthermore, we need to consider the propagation of experimental errors from the experimental
few-nucleon data that the 3NF contact terms are fitted to. Also this will be negligible as long as 
the 3NFs are adjusted to data with very small experimental errors; for example the empirical
binding energy of the triton is $8.481795 \pm 0.000002$ MeV, which will definitely lead to
negligible propagation. 

Now turning to the Hamiltoninan, we have to, first, account for uncertainties in the $NN$ and
$3N$ LECs due to the way they are fixed.
Based upon our experiences from Ref.~\cite{Mar13} and the fact that chiral EFT is
a low-energy expansion, we have fitted the $NN$ contact LECs to the $NN$ data below
100 MeV at LO and NLO, below 190 MeV at NNLO, and below 290 MeV at N$^3$LO and N$^4$LO.
One could think of choosing these fit-intervals slightly different and
a systematic investigation of the impact of such variation on the $NN$ LECs 
is still outstanding. However, we do not anticipate 
that large uncertainties would emerge from this source of error.

The story is different for the 3NF contact LECs, since several, 
very different procedures are in use for 
how to fix, e.~g., the two contact parameters of the NNLO 3NF, 
known as the $c_D$ and $c_E$ parameters (and once the ten 3NF contacts at N$^4$LO come into play, the situation will be even more divers).
Since, at NNLO, two parameters have to be fixed, two data are needed.
In most procedures, one of them is the triton binding energy.
For the second datum, the following choices have been made:
 the $nd$ doublet scattering length $^2a_{nd}$~\cite{Epe02},
the binding energy of $^4$He~\cite{Nog06},
the point charge radius radius of $^4$He~\cite{Heb11},
the Gamow-Teller matrix element of tritium $\beta$-decay~\cite{GP06,GQN09,Mar12}.
Alternatively,  the $c_D$ and $c_E$ parameters have also been pinned down by just
an optimal over-all fit of the properties of light nuclei~\cite{Nav07a}.
3NF contact LECs determined by different procedures will lead to different predictions
for the observables that were not involved in the fitting procedure. The differences in those results
establish the uncertainty.  
Specifically, it would be of interest to investigate the differences that occur
for the properties of intermediate-mass nuclei and nuclear matter when 3NF LECs fixed by different protocols
are applied.

The uncertainty in the $\pi N$ LECs used to be a large source of 
uncertainty, in particular, for predictions for many-body systems~\cite{Kru13,DHS16,Dri16}.
With the new, high-precision determination of the $\pi N$ LECs 
in the Roy-Steiner equations analysis~\cite{Hof15} (cf.\ Table~\ref{tab_lecs})
this large uncertainty is essentially eliminated, which is great progress, since it substantially reduces the error budget. We have varied the $\pi N$ LECs within the errors given in 
Table~\ref{tab_lecs} and find that the changes caused by these variations can easily be compensated by small readjustments of the $NN$ LECs resulting in essentially identical
phase shifts and $\chi^2$ for the fit of the data. Thus, this source of error is essentially negligible. The $\pi N$ LECs also appear in the 3NFs,
which also include contacts that can be used for readjustment. Future calculations of finite nuclei and nuclear matter should investigate what residual changes remain after such readjustment
(that would represent the uncertainty).
We expect this to be small.

The choice of the regulator function and its cutoff parameter create uncertainty.  
Originally, cutoff variations were perceived as a demonstration of the uncertainty at a given order
(equivalent to the truncation error).
However, in various investigations~\cite{Sam15,EKM15} it has been demonstrated that this is not correct and that cutoff variations,
in general, underestimate this uncertainty. 
Therefore, the truncation error is better determined by sticking literally to what
 `truncation error' means, namely, the error due to
 ignoring contributions from orders beyond the given order $\nu$. 
 The largest such contribution is the one of order $(\nu + 1)$,
 which one may, therefore, consider as representative for the magnitude of what is left out.
This suggests that the truncation error at order $\nu$ can reasonably be defined as
\begin{equation}
\Delta X_\nu = |X_\nu - X_{\nu+1}| \,, 
\end{equation}
where $X_\nu$ denotes the prediction for observable $X$ at order $\nu$. If $X_{\nu+1}$ is not available, then one may use, 
\begin{equation}
\Delta X_\nu = |X_{\nu-1} - X_\nu|Q/\Lambda \,,
\end{equation}
 choosing a typical value for the momentum $Q$, or $Q=m_\pi$.
Alternatively, one may also apply more elaborate definitions, like the one given in Ref.~\cite{EKM15}.
Note that one should not add up (in quadrature) the uncertainties due to regulator dependence and the truncation error, because they are not independent. In fact, it is appropriate to 
leave out the uncertainty due to regulator dependence entirely and just focus on the truncation error~\cite{EKM15}. The latter should be estimated using the same cutoff (e.~g., $\Lambda = 500$ MeV)
in all orders considered.

Finally, the last uncertainty to be taken into account is the uncertainty in the few- and many-body methods applied in the {\it ab inition} calculation.
This source of error has nothing to do with EFT.
Few-body problems are nowadays exactly solvable such that the error is negligible in those cases.
For heavier nuclei and nuclear matter, there are definitely uncertainties no matter what method is used. These uncertainties need to be estimated by the practitioners of those methods. But with the improvements of algoriths and the increase of computing power these errors are decreasing.

The bottom line is that the most substantial uncertainty is the truncation error. This is the dominant source of (systematic) error that should be carefully estimated for any calculation applying chiral 2NFs and 3NFs up to a given order.

\section{Summary and Conclusions}
\label{sec_concl}

We have constructed chiral $NN$ potentials through five orders of chiral EFT ranging from
LO to N$^4$LO~\cite{note2}. The construction may be perceived as consistent, because the same power 
counting scheme as well as the same cutoff procedures are applied in all orders.
Moreover, the long-range part of these potentials are fixed by the very accurate $\pi N$ LECs 
as determined in the Roy-Steiner equations analysis of Ref.~\cite{Hof15}. In fact,
the uncertainties of these LECs are so small that a variation within the errors leads to effects
that are essentially negligible at the current level of precision.
Another aspect that has to do with precision is that, at least at the highest order (N$^4$LO),
the $NN$ data below pion-production threshold are reproduced with the outstanding
$\chi^2$/datum of 1.15. This is the highest precision ever accomplished with any chiral
$NN$ potential to date.

The $NN$ potentials presented in this paper may serve as a
solid basis for systematic {\it ab initio} calculations of nuclear structure and reactions
that allow for a comprehensive error analysis. In particular, the order by order development of the potentials
will make possible a reliable determination of the truncation error at each 
order.

Our family of potentials is non-local and, generally, of soft character. This feature is reflected in the fact that
the predictions for the triton binding energy (from two-body forces only) converges to about 8.1 MeV
at the highest orders. This leaves room for moderate three-nucleon forces.

These features of our potentials are in contrast to other families of chiral $NN$ potentials
of local or semi-local character that have recently entered the market~\cite{Gez14,Pia15,Pia16,EKM15,EKM15a}.
Such potentials are less soft and, consequently, require stronger three-body force contributions.

The availability of families of chiral $NN$ potentials of different character offers
the opportunity for interesting systematic studies that
may ultimately shed light on issues, like,
the ``radius problem''~\cite{Lap16}, the overbinding of intermediate-mass nuclei~\cite{Bin14}, and others.

Note that the differences between the above-mentioned families of potentials are in the 
off-shell character,
which is not an observable. Thus, any off-shell behavior of a $NN$ potential is legitimate.
There is no wrong off-shell character.
However, some off-shell behaviors may lead in a more efficient way to realistic results than others.
That is of interest to the many-body practitioner. We are now in a position to systematically investigate this issue for chiral forces.

\acknowledgments
The work by R.M.\ and Y.N.\ was supported in part by the U.S. Department of Energy
under Grant No.~DE-FG02-03ER41270.
 The work by D.R.E.\ has been partially funded by MINECO under Contracts No.~FPA2013-47433-C2-2-P and FPA2016-77177-C2-2-P, and by the Junta de Castilla y Leon under Contract No.~SA041U16.

\appendix

\section{Phaseshift tables}
\label{app_ph}

In this appendix, we show the phase shifts as predicted by the N$^4$LO potential with $\Lambda=500$ MeV.
Note that our $pp$ phase shifts 
are the phase shifts of the nuclear plus relativistic Coulomb interaction with respect
to Coulomb wave functions.
For $nn$ and $np$ scattering, our phase shifts are the ones from the nuclear interaction
with respect to Riccati-Bessel functions. For more technical details of our phase shift calculations,
we refer the interested reader to the
appendix A3 of Ref.~\cite{Mac01}.

\begin{table}
\caption{$pp$ phase shifts (in degrees) up to $F$-waves at N$^4$LO ($\Lambda = 500$ MeV).}
\label{tab_pp}
\smallskip
\begin{tabular*}{\textwidth}{@{\extracolsep{\fill}}crrrrrrrrr}
\hline 
\hline 
\noalign{\smallskip}
 $T_{\rm lab}$ (MeV)
 & $^1S_0$
 & $^3P_0$
 & $^3P_1$
 & $^1D_2$
 & $^3P_2$
 & $^3F_2$
 & $\epsilon_2$
 & $^3F_3$
 & $^3F_4$
\\
 \hline 
     1 &  32.79 &   0.14 &  -0.08 &   0.00 &   0.01 &   0.00 &   0.00 &   0.00 &      0.00 \\
    5 &  54.84 &   1.61 &  -0.89 &   0.04 &   0.23 &   0.00 &  -0.05 &   0.00 &      0.00 \\
   10 &  55.20 &   3.79 &  -2.02 &   0.17 &   0.69 &   0.01 &  -0.20 &  -0.03 &      0.00 \\
   25 &  48.62 &   8.66 &  -4.84 &   0.69 &   2.57 &   0.11 &  -0.81 &  -0.23 &      0.02 \\
   50 &  38.84 &  11.42 &  -8.26 &   1.67 &   5.87 &   0.35 &  -1.69 &  -0.68 &      0.12 \\
  100 &  24.97 &   9.15 & -13.48 &   3.61 &  10.70 &   0.83 &  -2.62 &  -1.46 &      0.51 \\
  150 &  15.04 &   4.55 & -17.72 &   5.45 &  13.57 &   1.16 &  -2.83 &  -1.98 &      1.07 \\
  200 &   7.10 &  -0.47 & -21.39 &   7.22 &  15.54 &   1.20 &  -2.71 &  -2.31 &      1.67 \\
  250 &   0.11 &  -5.89 & -25.12 &   8.85 &  17.01 &   0.92 &  -2.42 &  -2.48 &      2.20 \\
  300 &  -6.43 & -11.40 & -29.35 &   9.91 &  17.84 &   0.35 &  -1.99 &  -2.46 &      2.59 \\
\hline
\hline
\noalign{\smallskip}
\end{tabular*}
\end{table}

\begin{table}
\caption{$nn$ phase shifts (in degrees) up to $F$-waves at N$^4$LO ($\Lambda = 500$ MeV).}
\label{tab_nn}
\smallskip
\begin{tabular*}{\textwidth}{@{\extracolsep{\fill}}crrrrrrrrr}
\hline 
\hline 
\noalign{\smallskip}
 $T_{\rm lab}$ (MeV)
 & $^1S_0$
 & $^3P_0$
 & $^3P_1$
 & $^1D_2$
 & $^3P_2$
 & $^3F_2$
 & $\epsilon_2$
 & $^3F_3$
 & $^3F_4$
\\
 \hline 
     1 &  57.62 &   0.21 &  -0.12 &   0.00 &   0.02 &   0.00 &   0.00 &   0.00 &      0.00 \\
    5 &  61.01 &   1.88 &  -1.03 &   0.05 &   0.28 &   0.00 &  -0.06 &  -0.01 &      0.00 \\
   10 &  57.82 &   4.16 &  -2.21 &   0.18 &   0.78 &   0.01 &  -0.22 &  -0.04 &      0.00 \\
   25 &  49.11 &   9.01 &  -5.08 &   0.73 &   2.77 &   0.11 &  -0.84 &  -0.24 &      0.02 \\
   50 &  38.71 &  11.55 &  -8.52 &   1.72 &   6.15 &   0.36 &  -1.72 &  -0.70 &      0.13 \\
  100 &  24.65 &   9.06 & -13.76 &   3.68 &  11.02 &   0.84 &  -2.62 &  -1.48 &      0.53 \\
  150 &  14.70 &   4.40 & -17.98 &   5.52 &  13.92 &   1.16 &  -2.82 &  -2.00 &      1.09 \\
  200 &   6.74 &  -0.63 & -21.62 &   7.28 &  15.94 &   1.20 &  -2.68 &  -2.32 &      1.70 \\
  250 &  -0.28 &  -6.02 & -25.32 &   8.88 &  17.42 &   0.91 &  -2.36 &  -2.49 &      2.23 \\
  300 &  -6.87 & -11.40 & -29.48 &   9.87 &  18.24 &   0.32 &  -1.93 &  -2.46 &      2.61 \\
\hline
\hline
\noalign{\smallskip}
\end{tabular*}
\end{table}

\begin{table}
\caption{$I=1$ $np$ phase shifts (in degrees) up to $F$-waves at N$^4$LO ($\Lambda = 500$ MeV).}
\label{tab_np1}
\smallskip
\begin{tabular*}{\textwidth}{@{\extracolsep{\fill}}crrrrrrrrr}
\hline 
\hline 
\noalign{\smallskip}
 $T_{\rm lab}$ (MeV)
 & $^1S_0$
 & $^3P_0$
 & $^3P_1$
 & $^1D_2$
 & $^3P_2$
 & $^3F_2$
 & $\epsilon_2$
 & $^3F_3$
 & $^3F_4$
\\
 \hline 
     1 &  62.00 &   0.18 &  -0.11 &   0.00 &   0.02 &   0.00 &   0.00 &   0.00 &      0.00 \\
    5 &  63.47 &   1.66 &  -0.92 &   0.04 &   0.27 &   0.00 &  -0.05 &   0.00 &      0.00 \\
   10 &  59.72 &   3.72 &  -2.03 &   0.16 &   0.75 &   0.01 &  -0.19 &  -0.03 &      0.00 \\
   25 &  50.48 &   8.25 &  -4.79 &   0.68 &   2.66 &   0.09 &  -0.76 &  -0.20 &      0.02 \\
   50 &  39.83 &  10.69 &  -8.20 &   1.68 &   5.96 &   0.31 &  -1.62 &  -0.61 &      0.11 \\
  100 &  25.68 &   8.25 & -13.44 &   3.68 &  10.76 &   0.78 &  -2.53 &  -1.35 &      0.49 \\
  150 &  15.78 &   3.63 & -17.67 &   5.56 &  13.63 &   1.08 &  -2.76 &  -1.86 &      1.04 \\
  200 &   7.90 &  -1.37 & -21.33 &   7.34 &  15.63 &   1.12 &  -2.64 &  -2.18 &      1.64 \\
  250 &   0.96 &  -6.75 & -25.05 &   8.96 &  17.12 &   0.83 &  -2.35 &  -2.35 &     2.17 \\
  300 &  -5.57 & -12.14 & -29.23 &   9.96 &  17.95 &   0.25 &  -1.93 &  -2.34 &     2.55 \\
\hline
\hline
\noalign{\smallskip}
\end{tabular*}
\end{table}

\begin{table}
\caption{$I=0$ $np$ phase shifts (in degrees) at N$^4$LO ($\Lambda = 500$ MeV).}
\label{tab_np0}
\smallskip
\begin{tabular*}{\textwidth}{@{\extracolsep{\fill}}crrrrrrrrr}
\hline 
\hline 
\noalign{\smallskip}
 $T_{\rm lab}$ (MeV)
 & $^1P_1$
 & $^3S_1$
 & $^3D_1$
 & $\epsilon_1$
 & $^3D_2$
 & $^1F_3$
 & $^3D_3$
 & $^3G_3$
 & $\epsilon_3$
\\
 \hline 
     1 &  -0.19 & 147.75 &  -0.01 &   0.11 &   0.01 &   0.00 &   0.00 &   0.00 &   0.00  \\
    5 &  -1.50 & 118.17 &  -0.19 &   0.68 &   0.22 &  -0.01 &   0.00 &   0.00 &   0.01  \\
   10 &  -3.06 & 102.61 &  -0.69 &   1.17 &   0.85 &  -0.07 &   0.00 &   0.00 &   0.08  \\
   25 &  -6.32 &  80.66 &  -2.83 &   1.79 &   3.71 &  -0.42 &   0.02 &  -0.05 &   0.56  \\
   50 &  -9.66 &  62.91 &  -6.48 &   2.03 &   8.82 &  -1.13 &   0.20 &  -0.26 &   1.62  \\
  100 & -14.78 &  43.72 & -12.20 &   2.09 &  16.51 &  -2.19 &   1.10 &  -0.94 &   3.54  \\
  150 & -19.52 &  31.42 & -16.34 &   2.33 &  21.08 &  -2.92 &   2.29 &  -1.76 &   4.95  \\
  200 & -23.46 &  21.60 & -19.55 &   2.99 &  23.89 &  -3.54 &   3.40 &  -2.57 &   5.90  \\
  250 & -25.72 &  12.68 & -22.01 &   4.09 &  25.21 &  -4.14 &   4.23 &  -3.24 &   6.40  \\
  300 & -25.27 &   4.02 & -23.38 &   5.34 &  24.41 &  -4.69 &   4.78 &  -3.65 &   6.39  \\
\hline
\hline
\noalign{\smallskip}
\end{tabular*}
\end{table}


\begin{thebibliography}{99}
\bibitem{ME11}
R. Machleidt and D. R. Entem, Phys. Rep. {\bf 503}, 1 (2011).
\bibitem{EHM09}
E. Epelbaum, H.-W. Hammer, and U.-G. Mei\ss ner,
Rev. Mod. Phys. {\bf 81}, 1773 (2009).
\bibitem{GL84}
J. Gasser and H. Leutwyler,
Ann. Phys. (N.Y.) {\bf 158}, 142 (1984).
\bibitem{GSS88}
J. Gasser, M. E. Sainio, and A. \v{S}varc,
Nucl. Phys. {\bf B307}, 779 (1988).
\bibitem{Wei90} S. Weinberg, Phys. Lett {\bf B251}, 288 (1990); Nucl.\ Phys.\ {\bf B363}, 3 (1991).
\bibitem{ORK94}
C. Ord\'o\~nez, L. Ray, and U. van Kolck,
Phys.\ Rev.\ Lett.\ {\bf 72}, 1982 (1994);
Phys.\ Rev.\ C {\bf 53}, 2086 (1996).
\bibitem{KBW97} N. Kaiser, R. Brockmann, and W. Weise,
Nucl.\ Phys.\ {\bf A625}, 758 (1997).
\bibitem{KGW98} N. Kaiser, S. Gerstend\"orfer, and W. Weise,
Nucl.\ Phys.\ {\bf A637}, 395 (1998).
\bibitem{Kai00a} N. Kaiser, Phys.\ Rev.\ C {\bf 61}, 014003 (2000).
\bibitem{Kai00b} N. Kaiser, Phys.\ Rev.\ C {\bf 62}, 024001 (2000).
\bibitem{Kai01} N. Kaiser, Phys.\ Rev.\ C {\bf 63}, 044010 (2001).
\bibitem{Kai01a} N. Kaiser, Phys.\ Rev.\ C {\bf 64}, 057001 (2001).
\bibitem{Kai02} N. Kaiser, Phys. Rev. C {\bf 65}, 017001 (2002).
\bibitem{EGM98} E. Epelbaum, W. Gl\"ockle, and U.-G. Mei\ss ner,
Nucl.\ Phys.\ {\bf A637}, 107 (1998); {\bf A671}, 295 (2000).
\bibitem{EM02} D. R. Entem and R. Machleidt,
Phys. Rev. C {\bf 66}, 014002 (2002).
\bibitem{EM03} D. R. Entem and R. Machleidt,
Phys. Rev. C {\bf 68}, 041001 (2003).
\bibitem{EGM05} E. Epelbaum, W. Gl\"ockle, and U.-G. Mei\ss ner,
Nucl. Phys. {\bf A747}, 362 (2005).
\bibitem{Nav07} 
  P.~Navratil,
  Few Body Syst.\  {\bf 41}, 117 (2007).
\bibitem{Eks13}
A.~Ekstr\H{o}m {\it et al.},
  Phys.\ Rev.\ Lett.\  {\bf 110},  192502 (2013).
\bibitem{Gez14}
A.~Gezerlis, I.~Tews, E.~Epelbaum, M.~Freunek, S.~Gandolfi, K.~Hebeler, A.~Nogga, and A.~Schwenk,
  Phys.\ Rev.\ C {\bf 90}, 054323 (2014).
  \bibitem{Pia15}
 M.~Piarulli, L.~Girlanda, R.~Schiavilla, R.~Navarro P\'{e}rez, J.~E.~Amaro, and E.~Ruiz Arriola,
  Phys.\ Rev.\ C {\bf 91},  024003 (2015).
  \bibitem{Pia16}
 M.~Piarulli, L.~Girlanda, R.~Schiavilla, A. Kievsky, A. Lovato, L. E. Marcucci, Steven C. Pieper, M. Viviani, and R. B. Wiringa,
  Phys.\ Rev.\ C {\bf 94},  054007 (2016).
  \bibitem{EKM15}
E. Epelbaum, H. Krebs, and Ulf-G. Mei\ss ner,
 Eur.\ Phys.\ J.\ A {\bf 51},  53 (2015).
 \bibitem{EKM15a}
E. Epelbaum, H. Krebs, and Ulf-G. Mei\ss ner,
 Phys. Rev. Lett. {\bf 115}, 122301 (2015).
\bibitem{PAA15}
 R.~Navarro P\'{e}rez, J.~E.~Amaro, and E.~Ruiz Arriola,
  Phys.\ Rev.\ C {\bf 91}, 054002 (2015), and more references to the {\it comprehensive} work by the Granada group therein. 
\bibitem{Eks15}
A.~Ekstr\H{o}m {\it et al.},
  Phys.\ Rev.\ C  {\bf 91},  051301 (2015).
\bibitem{Car16}
  B.~D.~Carlsson {\it et al.},
  Phys.\ Rev.\ X {\bf 6}, 011019 (2016).
  \bibitem{Tew16} 
  I.~Tews, S.~Gandolfi, A.~Gezerlis, and A.~Schwenk,
  Phys.\ Rev.\ C {\bf 93}, 024305 (2016).
  \bibitem{Lyn16} 
  J.~E.~Lynn, I.~Tews, J.~Carlson, S.~Gandolfi, A.~Gezerlis, K.~E.~Schmidt, and A.~Schwenk,
  Phys.\ Rev.\ Lett.\  {\bf 116}, 062501 (2016).
  
  \bibitem{Ren16}
Xiu-Lei Ren, Kai-Wen Li, Li-Sheng Geng, Bingwei Long, Peter Ring, and Jie Meng,
 Leading order covariant chiral nucleon-nucleon interaction,
 arXiv:1611.08475 [nucl-th].
  
  

  

\bibitem{Epe02}
E.~Epelbaum, A.~Nogga, W.~Gloeckle, H.~Kamada, U.~G.~Meissner, and H.~Witala,
  Phys.\ Rev.\ C {\bf 66}, 064001 (2002).
    \bibitem{NRQ10} 
  P.~Navratil, R.~Roth, and S.~Quaglioni,
  Phys.\ Rev.\ C {\bf 82}, 034609 (2010).
  
\bibitem{Viv13}
M. Viviani, L. Girlanda, A. Kievsky, and L. E. Marcucci, 
Phys. Rev. Lett. {\bf 111}, 172302 (2013).
\bibitem{Gol14}
J.~Golak {\it et al.},
  Eur.\ Phys.\ J.\ A {\bf 50}, 177 (2014)
  
  
  
  
  \bibitem{BNV13} 
  B.~R.~Barrett, P.~Navratil, and J.~P.~Vary,
  Prog.\ Part.\ Nucl.\ Phys.\  {\bf 69}, 131 (2013).
  
  \bibitem{Her13} 
  H.~Hergert, S.~K.~Bogner, S.~Binder, A.~Calci, J.~Langhammer, R.~Roth, and A.~Schwenk,
  Phys.\ Rev.\ C {\bf 87}, 034307 (2013).
  
  \bibitem{Hag14a} 
  G.~Hagen, T.~Papenbrock, M.~Hjorth-Jensen, and D.~J.~Dean,
  Rept.\ Prog.\ Phys.\  {\bf 77}, 096302 (2014).
  
  
  \bibitem{Sim17} 
  J.~Simonis, S.~R.~Stroberg, K.~Hebeler, J.~D.~Holt, and A.~Schwenk,
  ``Saturation with chiral interactions and consequences for finite nuclei,''
  arXiv:1704.02915 [nucl-th].
  
  \bibitem{HS10} 
  K.~Hebeler and A.~Schwenk,
  Phys.\ Rev.\ C {\bf 82}, 014314 (2010).
   \bibitem{Heb11} 
  K.~Hebeler, S. K. Bogner, R. J. Furnstahl, A. Nogga, and A.~Schwenk,
  Phys.\ Rev.\ C {\bf 83}, 031301(R) (2011).
  
  
  
  \bibitem{Hag14b} 
  G.~Hagen, T.~Papenbrock, A.~Ekstr\H{o}m, K.~A.~Wendt, G.~Baardsen, S.~Gandolfi, M.~Hjorth-Jensen, and C.~J.~Horowitz,
  Phys.\ Rev.\ C {\bf 89}, 014319 (2014).
  
  \bibitem{Cor13} 
L. Coraggio, J. W. Holt, N. Itaco, R. Machleidt, and F. Sammarruca,  
{ Phys. Rev.} C {\bf 87}, 014322 (2013).
  \bibitem{Cor14}
  L. Coraggio, J. W. Holt, N. Itaco, R. Machleidt, L. E. Marcucci, and F. Sammarruca,  
{ Phys. Rev.} C {\bf 89}, 044321 (2014).
  \bibitem{Sam15}
  F. Sammarruca, L. Coraggio, J. W. Holt, N. Itaco, R. Machleidt, and L. E. Marcucci,   
{ Phys. Rev.} C {\bf 91}, 054311 (2015).
  
  
  
    \bibitem{EMW02}
D. R. Entem, R. Machleidt, and H. Witala,
Phys. Rev. C {\bf 65}, 064005 (2002).
  
  

  
\bibitem{KGE12}
H. Krebs, A. Gasparyan, and E. Epelbaum, 
Phys. Rev. C {\bf 85}, 054006 (2012).
\bibitem{KGE13}
H. Krebs, A. Gasparyan, and E. Epelbaum, 
Phys. Rev. C {\bf 87}, 054007 (2013).  
\bibitem{Epe15}
E.~Epelbaum, A.~M.~Gasparyan, H.~Krebs, and C.~Schat,
  Eur.\ Phys.\ J.\ A {\bf 51},  26 (2015).
\bibitem{GKV11}
L. Girlanda, A. Kievsky, M. Viviani, 
Phys. Rev. C {\bf 84}, 014001 (2011).
\bibitem{Lap16} 
  V.~Lapoux, V.~Som\`{a}, C.~Barbieri, H.~Hergert, J.~D.~Holt, and S.~R.~Stroberg,
  Phys.\ Rev.\ Lett.\  {\bf 117}, 052501 (2016).
\bibitem{Bin14} 
  S.~Binder, J.~Langhammer, A.~Calci, and R.~Roth,
  Phys.\ Lett.\ B {\bf 736}, 119 (2014).


\bibitem{Ent15a}
D. R. Entem, N. Kaiser, R. Machleidt, and Y. Nosyk,
Phys. Rev. C {\bf 91}, 014002 (2015).
\bibitem{Ent15b}
D. R. Entem, N. Kaiser, R. Machleidt, and Y. Nosyk,
Phys. Rev. C {\bf 92}, 064001 (2015).
\bibitem{FPW15} 
R.~J. Furnstahl, D.~R. Phillips, and S. Wesolowski,
J. Phys. G {\bf 42}, 034028 (2015).
\bibitem{MWF17}
J. A. Melendez, S. Wesolowski, and R. J. Furnstahl,
Bayesian truncation errors in chiral effective field theory: nucleon-nucleon observables,
arXiv:1704.03308 [nucl-th], and references therein.


\bibitem{PDG}
K. A. Olive {\it et al.} (Particle Data Group), 
Chin. Phys. C {\bf 38}, 090001 (2014).
\bibitem{LM98b} G. Q. Li and R. Machleidt, Phys. Rev. C {\bf 58},
3153 (1998).
\bibitem{EGM04} E. Epelbaum, W. Gl\"ockle, and U.-G. Mei\ss ner,
Eur. Phys. J. A {\bf 19}, 125 (2004).
\bibitem{note1}
2015 Dr.\ Klaus Erkelenz Price Winners, University
of Bonn, Germany; for more information, see https://www.hiskp.uni-bonn.de

\bibitem{Hof15}
M. Hoferichter, J. Ruiz de Elvira, B. Kubis, and U.-G. Mei\ss ner,
Phys. Rev. Lett. {\bf 115}, 192301 (2015); Phys. Rep. {\bf 625}, 1 (2016).
\bibitem{AS83}
G. J. M. Austin and J. J. de Swart, Phys. Rev. Lett. {\bf 50}, 2039 (1983).
\bibitem{Ber88} J. R. Bergervoet, P. C. van Campen, W. A. van der Sanden,
and J. J. de Swart, Phys. Rev. C {\bf 38}, 15 (1988).
\bibitem{Kol98} 
U. van Kolck, M. C. M. Rentmeester, J. L. Friar, T. Goldman, and J. J. de Swart,
Phys. Rev. Lett. {\bf 80}, 4386 (1998).
\bibitem{BS66} R. Blankenbecler and R. Sugar, 
Phys.\ Rev.\/ {\bf 142}, 1051 (1966).

\bibitem{Mac01} 
R. Machleidt, Phys. Rev. C {\bf 63}, 024001 (2001).
\bibitem{EAH71} 
K. Erkelenz, R. Alzetta, and K. Holinde, Nucl. Phys. {\bf 
A176}, 413 (1971).
\bibitem{Mac93}
R. Machleidt, in: {\it Computational Nuclear Physics 2 -- Nuclear Reactions},
edited by K. Langanke, J.A. Maruhn, and S.E. Koonin (Springer, New York, 1993)
p.~1.

\bibitem{KSW96}
D. B. Kaplan, M. Savage, and M. B. Wise,   
{ Nucl. Phys.} {\bf B478}, 629 (1996);
{\ Phys. Lett.} {\it B424}, 390 (1998);
{\it Nucl. Phys.} {\bf B534}, 329 (1998).


\bibitem{Bir06}
M. C. Birse, 
{ Phys. Rev. C} {\bf 74}, 014003 (2006); 
{\it ibid.} {\bf 76}, 034002 (2007);
{\it ibid.} {\bf 77}, 047001 (2008).

\bibitem{LY12}
B. Long and C. Yang, 
Phys. Rev. C {\bf 86}, 024001 (2012).

\bibitem{Lon16}
B. Long, Int. J. Mod. Phys. E {\bf 25}, 1641006 (2016).


\bibitem{Val16}
M. P. Valderrama, Int. J. Mod. Phys. E {\bf 25}, 1641007 (2016).
\bibitem{Val16a}
M. P. Valderrama, M. Sanchez Sanchez, C.-J. Yang, Bingwei Long, J. Carbonell, and U. van Kolck,
Power Counting in Peripheral PartialWaves: The Singlet Channels,
arXiv:1611.10175.



\bibitem{San17}
M. Sanchez Sanchez, C.-J. Yang, Bingwei Long, and U. van Kolck,
The Two-Nucleon $^1S_0$ Amplitude Zero
in Chiral Effective Field Theory,
arXiv:1704.08524.

\bibitem{EGM17}
E. Epelbaum, J. Gegelia, and Ulf-G. Mei\ss ner,
Wilsonian renormalization group versus subtractive
renormalization in effective field theories
for nucleon–nucleon scattering,
arXiv:1705.02524.

\bibitem{Kon17} 
  S.~K\H{o}nig, H.~W.~Grie\ss hammer, H.-W.~Hammer, and U.~van Kolck,
  Phys.\ Rev.\ Lett.\  {\bf 118}, 202501 (2017).


\bibitem{Lep97}
G. P. Lepage,
How to Renormalize the Schr\"odinger Equation,
arXiv:nucl-th/9706029.
\bibitem{Mar13}
E. Marji {\it et al.}, Phys. Rev. C {\bf 88}, 054002 (2013).





\bibitem{Ber90}
J. R. Bergervoet, P. C. van Campen, R. A. M. Klomp, J.-L. de Kok, T. A. Rijken,
V. G. J. Stoks, and J. J. de Swart,
Phys. Rev. C {\bf 41}, 1435 (1990).
\bibitem{Sto93} 
V.\ G.\ J.\ Stoks, R.\ A.\ M.\ Klomp, 
M.\ C.\ M.\ Rentmeester, and J.\ J.\ de Swart, 
Phys.\ Rev.\ C {\bf 48}, 792 (1993).

\bibitem{GS08}
F. Gross and A. Stadler, Phys. Rev. C {\bf 78}, 014005 (2008).

\bibitem{PAA13}
 R.~Navarro P\'{e}rez, J.~E.~Amaro, and E.~Ruiz Arriola,
  Phys.\ Rev.\ C {\bf 88}, 064002 (2013),

\bibitem{Al04}
D. Albers {\it et al.}, Eur. Phys. J. A {\bf 22}, 125 (2004).

\bibitem{Hos68}
N. Hoshizaki, Prog. Theor. Phys. Suppl. {\bf 42}, 107 (1968).
\bibitem{BLW78}
J. Bystricky, F. Lehar, and P. Winterschnitz, J. Phys. (Paris) {\bf 39}, 1 (1978).

\bibitem{Ab01}
W. P. Abfalterer, F. B. Bateman, F. S. Dietrich, R. W. Finlay, R. C. Haight, and G. L. Morgan,
Phys. Rev. C {\bf 63}, 044608 (2001).
\bibitem{Bo02}
N. Boukharouba, F. B. Bateman, C. E. Brient, A. D. Carlson, S. M. Grimes, R. C. Haight,
T. N. Massey, and O. A. Wasson,
Phys. Rev. C {\bf 65}, 014004 (2002).
\bibitem{Me04}
P. Mermod {\it et al.}, Phys. Lett. B {\bf 597}, 243 (2004).
\bibitem{Me05}
P. Mermod {\it et al.}, Phys. Rev. C {\bf 72}, 061002 (2005).
\bibitem{Kl02}
J. Klug {\it et al.}, Nucl. Instrum. Methods A {\bf 489}, 282 (2002).
\bibitem{Bl04}
V. Blideanu {\it et al.}, Phys. Rev. C {\bf 70}, 014607 (2004).
\bibitem{Jo05}
C. Johansson {\it et al.}, Phys. Rev. C {\bf 71}, 024002 (2005).

\bibitem{Ar00a}
J. Arnold {\it et al.}, Eur. Phys. J. C {\bf 17}, 67 (2000).
\bibitem{Ar00b}
J. Arnold {\it et al.}, Eur. Phys. J. C {\bf 17}, 83 (2000).
\bibitem{Da02}
M. Daum {\it et al.}, Eur. Phys. J. C {\bf 25}, 55 (2002).

\bibitem{SS93}
V. Stoks and J. J. de Swart, Phys. Rev. C {\bf 47}, 761 (1993).
\bibitem{Sto94} 
V.\ G.\ J.\ Stoks, R.\ A.\ M.\ Klomp, 
C.\ P.\ F.\ Terheggen, and J.\ J.\ de Swart, 
Phys.\ Rev.\ C {\bf 49}, 2950 (1994).
\bibitem{WSS95} R.\ B.\ Wiringa, V.\ G.\ J.\ Stoks, and R. Schiavilla,
Phys.\ Rev.\ C {\bf 51}, 38 (1995).
\bibitem{MHE87}
R. Machleidt, K. Holinde, and Ch. Elster,
Phys. Rep. {\bf 149}, 1 (1987).

\bibitem{SP07} 
R. A. Arndt, W. J. Briscoe, I. I. Strakovsky, and R. L. Workman,
Phys. Rev. C {\bf 76}, 025209 (2007).

\bibitem{Sto95}
V. G. J. Stoks (private communication).

\bibitem{SES83} W. A. van der Sanden, A. H. Emmen, and J. J. de
Swart, Report No.\ THEF-NYM-83.11, Nijmegen (1983), unpublished;
quoted in Ref.~\cite{Ber88}.
\bibitem{Gon06} D. E. Gonz\'alez Trotter {\it et al.}, 
Phys. Rev. C {\bf 73}, 034001 (2006).
\bibitem{Che08} Q. Chen {\it et al.}, Phys. Rev. C {\bf 77}, 054002 (2008).



\bibitem{MNS90} G.\ A.\ Miller, M.\ K.\ Nefkens, and I.\ Slaus, Phys.\ Rep.\
{\bf 194}, 1 (1990).
\bibitem{Jen11}
U. D. Jentschura, A. Matveev, C. G. Parthey, J. Alnis, R. Pohl, Th. Udem, N. Kolachevsky, 
and T. W. H\H{a}nsch,
Phys. Rev. A {\bf 83}, 042505 (2011).
\bibitem{PG90}
W. N. Polyzou and W. Gl\H{o}ckle,
Few-Body Syst.\ {\bf 9}, 97 (1990).


\bibitem{Kol94} 
U. van Kolck, Phys. Rev. C {\bf 49}, 2932 (1994).
\bibitem{Ber08}
V. Bernard, E. Epelbaum, H. Krebs, and Ulf-G. Mei\ss ner,
Phys. Rev. C {\bf 77}, 064004 (2008).
\bibitem{Ber11}
V. Bernard, E. Epelbaum, H. Krebs, and Ulf-G. Mei\ss ner,
Phys. Rev. C {\bf 84}, 054001 (2011).
\bibitem{Heb15}
  K.~Hebeler, H.~Krebs, E.~Epelbaum, J.~Golak, and R.~Skibinski,
  Phys.\ Rev.\ C {\bf 91}, 044001 (2015).
\bibitem{Arn06} R. A. Arndt, W. J. Briscoe, I. I. Strakovsky, and R. L. Workman,
Phys. Rev. C {\bf 74}, 045205 (2006).
\bibitem{FM57}
J.-I. Fujita and H. Miyazawa, Prog. Theor. Phys. {\bf 17}, 360 (1957). 



\bibitem{PAA14} 
R. Navarro Perez, J. E. Amaro, and E. Ruiz Arriola, 
Phys. Rev. C {\bf 89}, 064006 (2014).
\bibitem{Per14} 
R. Navarro Perez, E. Garrido, J. E. Amaro, and E. Ruiz Arriola, 
Phys. Rev. C {\bf 90}, 047001 (2014).
\bibitem{Per15} 
R. Navarro Perez, J. E. Amaro, E. Ruiz Arriola, P. Maris, and J. P. Vary, 
Phys. Rev. C {\bf 92}, 064003 (2015).


\bibitem{Nog06}
A. Nogga, P. Navratil, B. R. Barrett, and J. P. Vary,
Phys.\ Rev.\ C {\bf 73} (2006) 064002.

\bibitem{GP06}
A. G\aa rdestig and D. R. Philips, Phys. Rev. Lett. {\bf 96} (2006) 232301.
\bibitem{GQN09}
D. Gazit, S. Quaglioni, and P. Navr\'atil, Phys. Rev. Lett. {\bf 103} (2009) 102502. 
\bibitem{Mar12} 
  L.~E.~Marcucci, A.~Kievsky, S.~Rosati, R.~Schiavilla, and M.~Viviani,
  Phys.\ Rev.\ Lett.\  {\bf 108}, 052502 (2012).
  
\bibitem{Nav07a}
P. Navratil, V. G. Gueorguiev, J. P. Vary, W. E. Ormand,
and A. Nogga, Phys. Rev. Lett. {\bf 99} (2007) 042501.

\bibitem{Kru13} 
  T.~Kr\H{u}ger, I.~Tews, K.~Hebeler, and A.~Schwenk,
  Phys.\ Rev.\ C {\bf 88}, 025802 (2013).
\bibitem{DHS16} 
  C.~Drischler, K.~Hebeler, and A.~Schwenk,
  Phys.\ Rev.\ C {\bf 93}, 054314 (2016).
\bibitem{Dri16} 
  C.~Drischler, A.~Carbone, K.~Hebeler, and A.~Schwenk,
  Phys.\ Rev.\ C {\bf 94}, 054307 (2016).


\bibitem{note2}
User-friendly FORTRAN codes for all $NN$ potentials presented in this paper
can be obtained from the authors upon request.





\end{thebibliography}
\end{document}